\documentclass[11pt]{article}
\usepackage{harvard}
\usepackage{amsmath}
\usepackage{amssymb}
\usepackage{graphicx}
\usepackage{theorem}
\usepackage{times}

\newcommand{\be}{\begin{equation}}
\newcommand{\ee}{\end{equation}}
\newcommand{\bea}{\begin{eqnarray}}
\newcommand{\eea}{\end{eqnarray}}
\newcommand{\Va}{\mbox{VaR}_\alpha}

\theoremheaderfont{\scshape}
\theoremstyle{break}
\theorembodyfont{\slshape}
\newtheorem{theorem}{Theorem}
\newtheorem{proposition}{Proposition}
\newtheorem{corollary}{Corollary}

\newtheorem{definition}{Definition}

\begin{document}

\title{VaR-Efficient Portfolios for a Class of Super- and Sub-Exponentially
Decaying Assets Return Distributions}

\author{Y. Malevergne$^{1,2}$ and D. Sornette$^{1,3}$ \\
\\
$^1$ Laboratoire de Physique de la Mati\`ere Condens\'ee CNRS UMR 6622\\
Universit\'e de Nice-Sophia Antipolis, 06108 Nice Cedex 2, France\\
$^2$ Institut de Science Financi\`ere et d'Assurances - Universit\'e Lyon I\\
43, Bd du 11 Novembre 1918, 69622 Villeurbanne Cedex\\
$^3$ Institute of Geophysics and Planetary Physics
and Department of Earth and Space Science\\
University of California, Los Angeles, California 90095, USA\\
\\
email: Yannick.Malevergne@unice.fr and sornette@unice.fr\\
fax: (33) 4 92 07 67 54\\
}

\date{}

\maketitle

\begin{abstract}
\noindent Using a family of modified Weibull distributions,
encompassing both sub-exponentials and super-ex\-po\-nen\-tials, to
parameterize the marginal distributions of asset returns and their
multivariate generalizations with Gaussian copulas, 
we offer exact formulas for the tails of the
distribution $P(S)$ of returns $S$ of a portfolio of arbitrary composition of these
assets. We find that the tail of $P(S)$ is also 
asymptotically a modified Weibull distribution
with a characteristic 
scale $\chi$ function of the asset weights with different functional forms
depending on the super- or sub-exponential behavior of the marginals
and on the strength of the dependence between the assets.
We then treat in details the problem of risk minimization using the
Value-at-Risk and Expected-Shortfall which are shown to be (asymptotically)
equivalent in this framework.

\end{abstract}

\section*{Introduction}

In recent years, the Value-at-Risk has become one of
the most popular risk assessment tool \cite{DP97,Jorion}. 
The infatuation for this particular risk
measure probably comes from a variety of factors,
the most prominent ones being its conceptual simplicity and relevance
in addressing the ubiquitous large risks often inadequately 
accounted for by the standard volatility,
and from its prominent role
in the recommendations of the international banking authorities
\cite{BIS}. Moreover, down-side risk measures 
such as the Value-at-risk seem more in
accordance with observed behavior of economic agents. For instance, according to
prospect theory \cite{KT79}, the perception of downward market movements is not
the same as upward movements. This may be reflected in the so-called
leverage effect, first 
discussed by \cite{Bl76}, who observed that the volatility
of a stock tends
to increase when its price drops (see \cite{Fouq,camplo,xu,BouMat}
for reviews and recent works).
Thus, it should be more natural to consider
down-side risk measures like the VaR than the variance traditionally used in
portfolio management \cite{Markovitz} which does
not differentiate between positive and negative change in future wealth.

However, the choice of the Value-at-Risk has recently been criticized
\cite{Szergo99,Danielsonetal01} due to its lack of coherence in the
sense of \citeasnoun{Artzner}, among other reasons. This deficiency
leads to several theoretical and practical problems. Indeed, other than
the class of elliptical distributions, the VaR is not sub-additive
\cite{Embrecht98}, and may lead to inefficient risk diversification
policies and to severe problems in the practical implementation
of portfolio optimization algorithms (see \cite{Chabaaneetal} for a
discussion). Alternative have been proposed in terms of
Conditional-VaR or Expected-Shortfall
\cite[for instance]{Artzner,AT02}, which enjoy the property of
sub-additivity. This ensures that they yield coherent portfolio allocations
which can be obtained by the simple linear optimization algorithm
proposed by \citeasnoun{Uryasev}.

From a practical standpoint, the estimation of the VaR of a portfolio is a
strenuous task, requiring large computational time leading sometimes to
disappointing results lacking accuracy and stability. As a consequence,
many approximation methods have
been proposed \cite[for instance]{TT01,EHJ02}. Empirical models constitute
another widely used approach,
since they provide a good trade-off between speed and
accuracy. 

From a general point of view, the parametric determination of the risks and
returns associated with a given portfolio constituted of $N$ assets is
completely embedded in the knowledge of their multivariate
distribution of returns. Indeed, the dependence between random
variables is completely described by their joint distribution. This
remark entails the two major problems of portfolio theory: 1) the
determination of the multivariate distribution function of asset returns; 2)
the derivation from it of a useful measure of portfolio risks, in the goal of
analyzing and optimizing portfolios. These objective can be easily reached
if one can
derive an analytical expression of the portfolio returns distribution from the
multivariate distribution of asset returns.

In the standard Gaussian framework, the multivariate distribution
takes the form of an exponential of minus a quadratic form $X'
\Omega^{-1} X$, where $X$ is the uni-column of asset returns and
$\Omega$ is their covariance matrix. The beauty and simplicity of the
Gaussian case is that the essentially impossible task of determining a
large multidimensional function is collapsed onto the very much
simpler one of calculating the $N(N+1)/2$ elements of the symmetric
covariance matrix. And, by the statibility of the Gaussian distribution,
the risk is then uniquely and completely embodied by
the variance of the portfolio return, which is easily determined from
the covariance matrix. This is the basis of \citeasnoun{Markovitz}'s portfolio
theory  and of the CAPM \cite{S64,L65,M66}. The same
phenomenon occurs in the stable Paretian portfolio analysis derived by
\cite{Fama65} and generalized
to separate positive and negative power law tails \cite{Bouchaudetal}. 
The stability of the distribution of returns is essentiel to bypass
the difficult problem of determining the
decision rules (utility function) of the economic agents
since all the risk measures are equivalent to a single parameter (the
variance in the case of a Gaussian universe).

However, it is well-known that the empirical distributions of
returns are neither Gaussian nor L\'evy Stable \cite{Lux,Gopikrishnan,GJ98} and
the dependences between assets are only imperfectly accounted for by
the covariance matrix \cite{Litterman}. It is thus desirable to find
alternative parameterizations of multivariate distributions
of returns which provide reasonably good
approximations of the asset returns distribution and which enjoy 
asymptotic stability properties in the tails so as to be relevant for the VaR.

To this aim, section 1 presents a specific parameterization
of the marginal distributions in terms of so-called modified Weibull
distributions introduced by \citeasnoun{Sornette1}, which are
essentially exponential of minus a power law. This 
family of distributions contains both sub-exponential and super-exponentials,
including the Gaussian law as a special case. It is shown that this
parameterization is relevant for modeling the distribution of asset returns
in both an unconditional and a conditional framework. The dependence structure
between the asset is described by a Gaussian copula which allows us to describe
several degrees of dependence: from independence to comonotonicity. The
relevance of the Gaussian copula has been put in light by several recent
studies \cite{SorAnderSim2,Sornette1,MS01,MS_mom}.

In section 2, we use the multivariate construction based on (i) the modified
Weibull marginal distributions and (ii) the Gaussian copula to derive the
asymptotic analytical form of the tail of the distribution of returns of a
portfolio composed of an arbitrary combination of these assets. 
In the case where individual asset returns have modified-Weibull
distributions, we show that the tail of the distribution of portfolio returns $S$ is
asymptotically of the same form but with a characteristic 
scale $\chi$ function of the asset weights
taking different functional forms
depending on the super- or sub-exponential behavior of the marginals
and on the strength of the dependence between the assets. 
Thus, this particular class of modified-Weibull distributions enjoys
(asymptotically) the same stability properties as the Gaussian or L\'evy distributions.
The dependence properties are shown to be 
embodied in the $N(N+1)/2$ elements of a
non-linear covariance matrix and the individual risk of each assets are
quantified by the sub- or super-exponential behavior of the marginals.

Section 3 then uses this
non-Gaussian nonlinear dependence framework
to estimate the Value-at-Risk (VaR) and the Expected-Shortfall. 
As in the
Gaussian framework, the VaR and the Expected-Shortfall are (asymptotically)
controlled only by the non-linear covariance matrix, leading to their
equivalence. More generally, any risk measure
based on the (sufficiently far) tail of the distribution of the portfolio returns are
equivalent since they can be expressed as a function of the non-linear
covariance matrix and the weights of the assets only.

Section 4 uses this set of results to offer an approach to portfolio
optimization based on the asymptotic form of the
tail of the distribution of portfolio returns. When possible, we give the
analytical formulas of the
explicit composition of the optimal portfolio or suggest the use of reliable
algorithms when numerical calculation is needed.

Section 5 concludes.

Before proceeding with the presentation of our results, we set the notations to
derive the basic problem addressed in this paper, namely to study the
distribution
of the sum of weighted random variables with given marginal
distributions and
dependence.  Consider a portfolio with $n_i$ shares of asset
$i$ of price $p_i(0)$ at time $t=0$ whose initial wealth is
\be
W(0) = \sum_{i=1}^N n_i p_i(0)~.
\ee
A time $\tau$ later, the wealth has become $W(\tau) = \sum_{i=1}^N n_i
p_i(\tau)$ and the wealth variation is
\be
\delta_{\tau} W \equiv W(\tau) -W(0) = \sum_{i=1}^N n_i p_i(0)
\frac{p_i(\tau) - p_i(0)}{p_i(0)}
= W(0) ~\sum_{i=1}^N w_i x_i(t,\tau) ,
\ee
where
\be
w_i = \frac{n_i p_i(0)}{\sum_{j=1}^{N} n_j p_j(0)}~
\ee
is the fraction in capital invested in the $i$th asset at time $0$ and
the return
$x_i(t,\tau)$ between time $t-\tau$ and $t$ of asset $i$ is defined as:
\be
 x_i(t,\tau) = \frac{p_i(t)-p_i(t-\tau)}{p_i(t-\tau)} ~.
 \label{jhghss}
\ee
Using the definition (\ref{jhghss}),
this justifies us to write the return $S_{\tau}$ of the portfolio over a time
interval $\tau$ as the weighted sum of the returns $r_i(\tau)$ of the assets
$i=1,...,N$ over the time interval $\tau$
\be
S{\tau} = \frac{\delta_{\tau} W}{W(0)} = \sum_{i=1}^N w_i ~x_i(\tau)~.
\label{jjkmmq}
\ee
In the sequel, we shall thus consider the asset returns $X_i$ as the
fundamental variables and study their aggregation properties, namely how the
distribution of portfolio
return equal to their weighted sum derives for their multivariable
distribution.
We shall consider a single time scale $\tau$ which can be chosen arbitrarily,
say equal to one day. We shall thus drop the dependence on $\tau$,
understanding
implicitly that all our results hold for returns estimated over time
step $\tau$.

\section{Definitions and important concepts}
\subsection{The modified Weibull distributions}

We will consider a class of distributions with fat tails but decaying faster
than any power law. Such possible behavior for assets returns distributions
have been suggested to be relevant by several empirical works
\cite{MS95,GJ98,MPS02} and has also been asserted to provide a convenient and
flexible parameterization of many phenomena found in nature and in the social
sciences \cite{Laherrere}.  In all the following, we will use the
parameterization introduced by \citeasnoun{Sornette1} and define the
modified-Weibull distributions:

\begin{definition}[Modified Weibull Distribution]
\label{def:mw} A random variable $X$ will be said to follow a modified Weibull
distribution with exponent $c$ and scale parameter $\chi$, denoted in the
sequel $X \sim {\cal W}(c, \chi)$, if and only if the random variable
\be
Y= \mbox{sgn}(X)~\sqrt{2} \left(\frac{|X|}{\chi} \right)^\frac{c}{2}
\label{gaufms}
\ee
follows a Normal distribution.$\square$
\end{definition}

These so-called modified-Weibull distributions
can be seen to be general forms of the extreme tails of product of
random variables
\cite{FS97}, and using the theorem of change of variable, we can assert that
the density of such distributions is
\be
\label{eq:weibull}
p(x)=\frac{1}{2\sqrt{\pi}} \frac{c}{\chi^{\frac{c}{2}}}
|x|^{\frac{c}{2}-1} e^{-\left( \frac{|x|}{\chi}\right)^c} \; ,
\ee
where $c$ and $\chi$ are the two key parameters.

These expressions are close to the
Weibull distribution, with the addition of a power law prefactor to the
exponential such that the Gaussian law is retrieved for $c=2$.
Following \citeasnoun{Sornette1}, \citeasnoun{SorAnderSim2} and
\citeasnoun{AndersenRisk}, we call (\ref{eq:weibull}) the modified Weibull
distribution. For $c<1$, the pdf is a stretched exponential, which belongs
to the class of sub-exponential.
The exponent $c$ determines the shape of the
distribution, fatter than an exponential if $c<1$. The parameter $\chi$
controls the scale or characteristic width of the distribution. It plays
a role analogous to the standard deviation of the Gaussian law.

The interest of these family of distributions for financial
purposes have also been recently underlined by  \citeasnoun{BG00} and
\citeasnoun{BGL}. Indeed these authors have shown that given a series of
return $\{r_t\}_t$ following a GARCH(1,1) process, the large deviations of the
returns $r_{t+k}$ and of the aggregated returns $r_t + \cdots + r_{t+k}$
conditional on the return at time $t$ are distributed according to a
modified-Weibull distribution, where the exponent $c$ is related to the number
of step forward $k$ by the formula $c=2/k$ .

A more general parameterization taking
into account a possible asymmetry between negative and positive
values (thus leading to possible non-zero mean) is
\bea
\label{eq:pasym1}
p(x)&=&\frac{1}{2\sqrt{\pi}} \frac{c_+}{\chi_+^{\frac{c_+}{2}}}
|x|^{\frac{c_+}{2}-1} e^{-\left(
\frac{|x|}{\chi_+}\right)^{c_+}}~~~~\mbox{if}~x\ge 0\\
\label{eq:pasym2}
p(x)&=&\frac{1}{2\sqrt{\pi}} \frac{c_-}{\chi_-^{\frac{c_-}{2}}}
|x|^{\frac{c_-}{2}-1} e^{-\left(
\frac{|x|}{\chi_-}\right)^{c_-}}~~~~\mbox{if}~x<0~.
\eea

\begin{figure}
\begin{center}
\includegraphics[width=15cm]{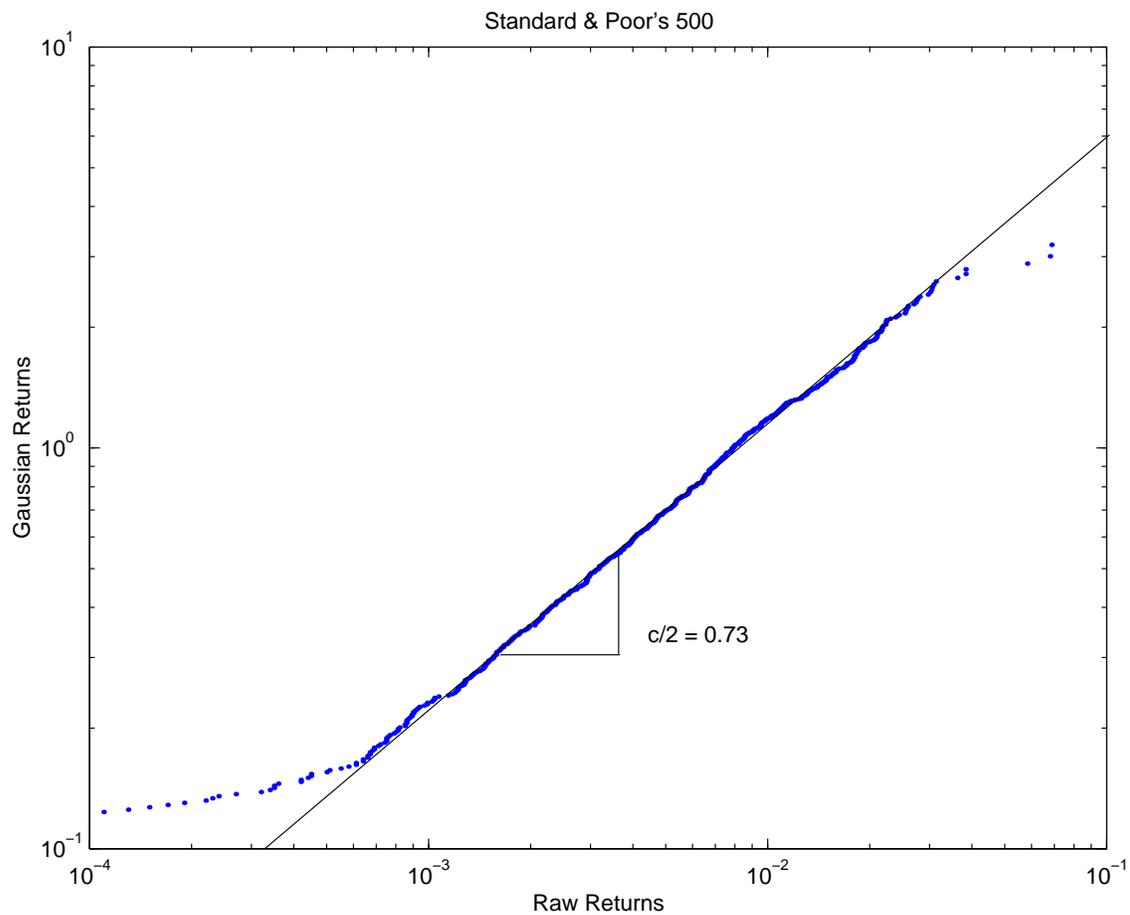}
\end{center}
\caption{\label{fig:sp500} Graph of the Gaussianized Standard \& Poor's 500
index returns versus its raw returns, during the time interval from January 03,
1995 to December 29, 2000 for the negative tail of the distribution.}
\end{figure}

In what follows, we will assume that the marginal probability distributions of
returns follow modified Weibull distributions. 
Figure \ref{fig:sp500} shows the (negative)
``Gaussianized'' returns $Y$ defined in (\ref{gaufms})
of the Standard and Poor's 500 index versus the
raw returns $X$ over the time interval from January 03, 1995 to December 29, 2000.
With such a representation, the modified-Weibull distributions are qualified by
a power law of exponent $c/2$, by definition~\ref{def:mw}. The double
logarithmic scales
of figure~\ref{fig:sp500} clearly shows a straight line over an extended
range of data, qualifying a power law relationship.
An accurate determination of the parameters $(\chi,c)$ can
be performed by maximum likelihood estimation \cite[pp 160-162]{S2000}.
However, note that, in the tail, the six most
extreme points significantly deviate from the modified-Weibull description.
Such an anomalous behavior of the most extreme returns can be 
probably be associated with the
notion of ``outliers'' introduced by \citename{JS98} \citeyear{JS98,JS02} and
associated with behavioral and crowd phenomena during turbulent market phases.

The modified Weibull distributions defined here are of interest for financial purposes
and specifically for portfolio and risk management, since they offer a flexible
parametric representation of asset returns distribution either in a conditional
or an unconditional framework, depending on the standpoint prefered by manager.
The rest of the paper uses this family of distributions.

\subsection{Tail equivalence for distribution functions}

An interesting feature of the modified Weibull distributions, as we will see in
the next section, is to enjoy the property of asymptotic stability. 
Asymptotic stability means that, in the regime of large deviations, a sum of
independent and identically distributed modified Weibull variables follows the
same modified Weibull distribution, up to a rescaling.

\begin{definition}[Tail equivalence]
\label{def:te}
Let $X$ and $Y$ be two random variables with distribution function $F$ and $G$
respectively.

$X$ and $Y$ are said to be equivalent in the upper tail if and only if there
exists $\lambda_+ \in (0, \infty)$ such that
\be
\lim_{x \rightarrow +\infty} \frac{1- F(x)}{1- G(x)}= \lambda_+.
\ee
Similarly, $X$ and $Y$ are said equivalent in the lower tail if
and only if there exists $\lambda_- \in (0, \infty)$ such that
\be
\lim_{x \rightarrow -\infty} \frac{F(x)}{G(x)}= \lambda_-.
\ee
$ \square$
\end{definition}

Applying l'Hospital's rule, this gives immediately the following corollary:

\begin{corollary}
\label{co:te}
Let $X$ and $Y$ be two random variables with densities functions $f$
and $g$ respectively.
$X$ and $Y$ are equivalent in the upper (lower) tail if and only if
\be
\lim_{x \rightarrow \pm \infty} \frac{f(x)}{g(x)}= \lambda_\pm,~ ~ ~
\lambda_\pm \in (0, \infty). \ee
$\square$
\end{corollary}

\subsection{The Gaussian copula}

We recall only the basic properties about copulas and refer the interested
reader to \cite{Nelsen}, for instance, for more information. 
Let us first give the definition of a copula
of $n$ random variables.
\begin{definition}[Copula]
A function  $C$ : $[0,1]^n \longrightarrow [0,1]$ is a $n$-copula if it
enjoys the following properties~:
\begin{itemize}
\item $\forall u \in [0,1] $, $C(1,\cdots, 1, u, 1 \cdots, 1)=u$~,
\item $\forall u_i \in [0,1] $, $C(u_1, \cdots, u_n)=0$ if
at least one of the $u_i$ equals zero~,
\item $C$ is grounded and $n$-increasing, i.e., the $C$-volume of every
boxes whose vertices lie in $[0,1]^n$ is positive.  $\square$
\end{itemize}
\end{definition}

The fact that such copulas can be very useful for representing
multivariate distributions with arbitrary marginals is seen from the
following result.
\begin{theorem}[Sklar's Theorem]
Given an $n$-dimensional
distribution function  $F$ with {\it continuous} marginal distributions
$F_1, \cdots, F_n$, there
exists a {\it unique} $n$-copula $C$ : $[0,1]^n \longrightarrow [0,1]$ such
that~: \be
F(x_1, \cdots, x_n) = C(F_1(x_1), \cdots, F_n(x_n))~.
\ee
$\square$
\end{theorem}

This theorem provides both a parameterization of
multivariate distributions and a construction scheme for copulas.
Indeed, given
a multivariate distribution $F$ with margins $F_1, \cdots, F_n$, the function
\be
\label{eq:cop}
C(u_1, \cdots, u_n)= F\left( F_1^{-1}(u_1), \cdots, F_n^{-1}(u_n) \right)
\ee
is automatically a $n$-copula. Applying this theorem to the multivariate
Gaussian distribution, we can derive the so-called {\it Gaussian copula}.

\begin{definition}[Gaussian copula]
 Let $\Phi$ denote the standard Normal distribution and
$\Phi_{V,n}$ the $n$-dimensional Gaussian distribution with correlation
matrix ${\bf V}$.  Then, the Gaussian $n$-copula
with correlation matrix ${\bf V}$ is \be
\label{eq:gc}
C_V(u_1, \cdots, u_n) = \Phi_{V,n} \left(\Phi^{-1}(u_1),
\cdots, \Phi^{-1}(u_n) \right)~,
\ee
whose density
\be
c_V(u_1, \cdots, u_n) = \frac{\partial C_V(u_1, \cdots,
u_n)}{\partial u_1 \cdots \partial u_n}
\ee
reads
\be
\label{eq:gcd}
c_V(u_1, \cdots, u_n) = \frac{1}{\sqrt{\det {\bf V}}} \exp \left( -
\frac{1}{2} y_{(u)}^t
({\bf V}^{-1}- \mbox{Id}) y_{(u)} \right)
\ee
with $y_k(u)=\Phi^{-1}(u_k)$.  Note that theorem 1 and equation
(\ref{eq:cop}) ensure that $C_V(u_1, \cdots, u_n)$ in equation (\ref{eq:gc})
is a copula. $\square$
\end{definition}

It can be shown that the Gaussian copula naturally arises when one tries to
determine the dependence between random variables using the principle of
entropy maximization  \cite[for instance]{Rao,Sornette1}. Its pertinence and
limitations for modeling the dependence between assets returns has been tested
by \citeasnoun{MS01}, who show that in most cases, this description of the
dependence can be considered satisfying, specially for stocks, provided
that one does not consider too extreme realizations \cite{MS_risk,MS_ext,zeevi}.

\section{Portfolio wealth distribution for several dependence structures}

\subsection{Portfolio wealth distribution for independent assets}

Let us first consider the case of a portfolio made of independent assets. This
limiting (and unrealistic) case is a mathematical idealization which provides
a first natural benchmark of the class of portolio return
distributions to be expected.
Moreover, it is generally the only case for which the calculations are
analytically tractable.  For such independepent assets distributed with the
modified Weibull distributions, the following results prove the
asymptotic stability of this set of distributions:

\begin{theorem}[Tail equivalence for i.i.d modified Weibull random variables]
\label{th:as}
Let $X_1, X_2, \cdots, X_N$ be $N$ independent and
identically ${\cal W}(c, \chi)$-distributed random
variables. Then, the variable
\be
S_N=X_1+X_2+ \cdots+X_N
\ee
is equivalent in the lower and upper tail to $Z \sim {\cal W}(c, \hat \chi)$,
with
\bea
\hat \chi &=& N^\frac{c-1}{c}\chi, ~~~ c>1, \\
\hat \chi &=& \chi, ~~~~~~~~~~~~~c \le 1.
\eea
$\square$
\end{theorem}

This theorem is a direct consequence of the theorem stated below and is based
on the result given by \citeasnoun{FS97} for $c>1$ and on general properties of
sub-exponential distributions when $c \le1$.

\begin{theorem}[Tail equivalence for weighted sums of independent variables]
\label{th:asws}
Let $X_1, X_2, \cdots, X_N$ be $N$ independent and
identically ${\cal W}(c, \chi)$-distributed random
variables.
Let $w_1$, $w_2, \cdots, w_N$ be $N$ non-random real coefficients.
Then, the variable
\be
S_N=w_1 X_1+w_2 X_2+ \cdots+w_N X_N
\ee
is equivalent in the upper and the lower tail to $Z \sim{\cal W}(c, \hat
\chi)$ with
\bea
\label{eq:ichi}
\hat \chi &=&\left( \sum_{i=1}^N |w_i|^\frac{c}{c-1} \right)^\frac{c-1}{c}
\cdot \chi,~~~~~~~~~~ c>1, \\
\hat \chi &=& \max_i\{|w_1|, |w_2|, \cdots, |w_N|\}, ~~~ c \le1.
\eea
$\square$
\end{theorem}

The proof of this theorem is given in appendix \ref{app:asws}.

\begin{corollary}
\label{co:1}
Let $X_1, X_2, \cdots, X_N$ be $N$ independent random variables such that
$X_i \sim{ \cal W}(c, \chi_i)$. Let $w_1$, $w_2, \cdots, w_N$ be $N$
non-random real coefficients. Then, the variable
\be
S_N=w_1 X_1+w_2 X_2+ \cdots+w_N X_N
\ee
is equivalent in the upper and the lower tail to $Z \sim{\cal W}(c, \hat
\chi)$ with
\bea
\label{eq:chi_sum_ind}
\hat \chi &=& \left( \sum_{i=1}^N |w_i \chi_i|^\frac{c}{c-1} \right)^\frac{c-1}{c}
~, ~~~c>1,\\
\hat \chi &=& \max_i\{|w_1 \chi_1|, |w_2 \chi_2|, \cdots, |w_N \chi_N|\}, ~~~ c
\le1. \eea
$\square$
\end{corollary}

The proof of the corollary is a straightforward application of theorem
\ref{th:asws}. Indeed, let $Y_1, Y_2, \cdots, Y_N$ be $N$ independent and
identically ${\cal W}(c, 1)$-distributed random variables. Then,
\be
(X_1, X_2, \cdots, X_N) \stackrel{d}{=} (\chi_1 Y_1, \chi_2 Y_2, \cdots,
\chi_N Y_N),
\ee
which yields
\be
S_N \stackrel{d}{=} w_1 \chi_1\cdot Y_1 + w_2 \chi_2 \cdot Y_2 + \cdots + w_N
\chi_N \cdot Y_N~.
\ee
Thus, applying theorem \ref{th:asws} to the i.i.d variables $Y_i$'s with weights
$w_i \chi_i$ leads to corollary \ref{co:1}.

\subsection{Portfolio wealth distribution for comonotonic assets}

The case of comonotonic assets is of interest as
the limiting case of the strongest possible dependence
between random variables. By definition,
\begin{definition}[Comonotonicity]
the variables $X_1, X_2, \cdots, X_N$ are comonotonic if and only
if there exits a random variable $U$ and non-decreasing functions $f_1, f_2,
\cdots, f_N$ such that
\be
(X_1, X_2, \cdots, X_N) \stackrel{d}{=}(f_1(U), f_2(U), \cdots, f_N(U)).
\ee
$\square$
\end{definition}
In terms of copulas, the comonotonicity can be
expressed by the following form of the copula
\be
C(u_1, u_2, \cdots, u_N) = \min(u_1,u_2,\cdots, u_N)~.
\ee
This expression is known as the Fr\'{e}chet-Hoeffding upper bound for copulas
\cite[for instance]{Nelsen}. It would be appealing to think that
estimating the Value-at-Risk under the comonotonicity assumption could provide
an upper bound for the Value-at-Risk. However, it turns out to be wrong, due
--as we shall see in the sequel-- to the lack of coherence (in the sense of
\citeasnoun{Artzner}) of the Value-at-Risk, in the general case.
Notwithstanding, an upper and lower bound can always be derived for the
Value-at-Risk \cite{EHJ02}. But in the present situation, where we are only
interested in the class of modified Weibull distributions with a Gaussian
copula, the VaR derived under the comonoticity assumption will actually
represent the upper bound (at least for the VaR calculated at sufficiently
hight confidence levels).

\begin{theorem}[Tail equivalence for a sum of comonotonic random variables]
\label{th:comonotony}
Let $X_1, X_2, \cdots, X_N$ be $N$ comonotonic random variables such that
$X_i \sim{ \cal W}(c, \chi_i)$. Let $w_1, w_2, \cdots, w_N$ be $N$
non-random real coefficients. Then, the variable
\be
S_N=w_1 X_1+w_2 X_2+ \cdots+w_N X_N
\ee
is equivalent in the upper and the lower tail to $Z \sim{\cal W}(c, \hat
\chi)$ with
\be
\hat \chi = \sum_i w_i \chi_i~.
\ee
$\square$
\end{theorem}

The proof is obvious since, under the assumption of comonotonicity, the
portfolio wealth $S$ is given by 
\be
S= \sum_i w_i \cdot X_i \stackrel{d}{=} \sum_{i=1}^N w_i \cdot f_i(U),
\ee
and for modified Weibull distributions, we have
\be
f_i(\cdot)=\mbox{sgn}(\cdot)~ \chi_i
\left(\frac{|\cdot|}{\sqrt{2}}\right)^{2/c_i},
\ee
in the symmetric case while $U$ is a Gaussian random variable. If, in
addition, we assume that all assets have the same exponent $c_i=c$, it
is clear that $S \sim {\cal W}(c, \hat \chi)$ with 
\be
\hat \chi = \sum_i w_i \chi_i.
\ee
It is important to note that this relation is exact and not asymptotic as in
the case of independent variables.

When the exponents $c_i$'s are different from an asset to another,
a similar result holds, since we can
still write the inverse cumulative function of $S$ as
\be
F_S^{-1}(p) = \sum_{i=1}^N w_i F_{X_i}^{-1} (p), ~~~~~p \in (0,1),
\ee
which is the property of {\it additive comonotonicity} of the
Value-at-Risk\footnote{This relation shows that, in general, the VaR
  calculated for comonotonic assets does not provide an upper bound of
  the VaR, whatever the dependence structure the portfolio may be. Indeed, in
such a case, we have ${\rm VaR}(X_1+X_2) = {\rm VaR}(X_1) + {\rm VaR}(X_2)$
while, by lack of coherence, we may have
${\rm VaR}(X_1+X_2) \ge {\rm VaR}(X_1) + {\rm VaR}(X_2)$ for some dependence
structure between $X_1$ and $X_2$.}. Let us then sort the $X_i$'s such that
$c_1 = c_2 = \cdots = c_p < c_{p+1} \le \cdots \le c_N$. We immediately obtain
that $S$ is equivalent in the tail to $Z \sim {\cal W}(c_1, \hat \chi)$, where
\be
\hat \chi = \sum_{i=1}^p w_i \chi_i.
\ee
In such a case, only the assets with the fatest tails contributes to
the behavior of the sum in the large deviation regime.

\subsection{Portfolio wealth under the Gaussian copula hypothesis}

\subsubsection{Derivation of the multivariate distribution with a Gaussian
copula and modified Weibull margins}

An advantage of the class of modified Weibull distributions
(\ref{eq:weibull}) 
is that the transformation into a Gaussian, and thus the calculation of the
vector ${\bf y}$ introduced in definition~\ref{def:mw}, is
particularly simple. It takes the form
\be
 \label{eq:yx}
y_k=\mbox{sgn}(x_k)~\sqrt{2}~\left(\frac{
|x_k|}{\chi_k}\right)^{\frac{c_k}{2}},
\ee
where $y_k$ is normally distributed . These variables $Y_i$ then allow us to
obtain the covariance matrix ${\bf
V}$ of the Gaussian copula : \be
V_{ij}=2 \cdot {\rm E} \left[ \mbox{sgn}(x_ix_j) \left(
\frac{|x_i|}{\chi_i} \right)^{\frac{c_i}{2}} \left( \frac{|x_j|}{\chi_j}
\right)^{\frac{c_j}{2}} \right] \; ,
\ee
which always exists and can be efficiently estimated. 
The multivariate density $P({\bf x})$ is thus given by:
\bea
P(x_1, \cdots , x_N) &=& c_V(x_1, x_2, \cdots, x_N)~ \prod_{i=1}^N p_i(x_i)\\
\label{eq:px}
&=& \frac{1}{2^N \pi^{N/2}\sqrt{V}} \prod_{i=1}^N
\frac{c_i |x_i|^{c/2-1}}{\chi_i^{c/2}} \exp\left[-\sum_{i,j}
V_{ij}^{-1}\left(\frac{|x_i|}{\chi_i}\right)^{c/2}
\left(\frac{|x_j|}{\chi_j}\right)^{c/2} \right]~.
\eea
Obviously, similar transforms hold, {\it mutatis mutandis}, for the
asymmetric case (\ref{eq:pasym1},\ref{eq:pasym2}).

\subsubsection{Asymptotic distribution of a sum of modified Weibull
  variables with the same exponent $c>1$}

We now consider a portfolio made of dependent assets with pdf given by
equation (\ref{eq:px}) or its asymmetric generalization. For such distributions of
asset returns, we obtain the following result

\begin{theorem}[Tail equivalence for a sum of dependent random variables]
\label{th:gauss}
Let $X_1, X_2, \cdots, X_N$ be $N$ random variables with a
dependence structure described by the Gaussian copula with correlation
matrix ${\bf V}$ and such that each $X_i \sim{ \cal W}(c,
\chi_i)$. Let $w_1, w_2, \cdots, w_N$ be $N$ (positive)
non-random real coefficients. Then, the variable
\be
S_N=w_1 X_1+w_2 X_2+ \cdots+w_N X_N
\ee
is equivalent in the upper and the lower tail to $Z \sim{\cal W}(c, \hat
\chi)$ with
\be
\label{eq:gchi}
\hat \chi = \left(\sum_i w_i \chi_i \sigma_i \right)^\frac{c-1}{c},
\ee
where the $\sigma_i$'s are the unique (positive) solution of
\be
\label{eq:eqsigma}
\sum_{i} V_{ik}^{-1} {\sigma_i}^{c/2}= w_k \chi_k ~ \sigma_k^{1-c/2}~,
~~~~~~~~~~~~\forall k~.
\ee
$\square$
\end{theorem}
The proof of this theorem follows the same lines as the proof of
theorem \ref{th:asws}. We thus only provide a heuristic derivation
of this result in appendix \ref{app:A}.
Equation (\ref{eq:eqsigma}) is equivalent to
\be
\label{eq:equiv_sys}
\sum_{k} V_{ik}~w_k \chi_k~{\sigma_k}^{1-c/2} =  {\sigma_i}^{c/2}~,
~~~~~~~~~~~~\forall i~.
\ee
which seems more attractive since it does not require the inversion of
the correlation matrix. In the special case where ${\bf V}$ is the
identity matrix, the variables
$X_i$'s are independent so that equation (\ref{eq:gchi}) must yield the same
result as equation (\ref{eq:ichi}). This results from the expression
of $\sigma_k= ( w_k \chi_k)^\frac{1}{c-1}$ valid in the
independent case. Moreover, in the limit
where all entries of {\bf V} equal one, we retrieve the case of comonotonic
assets. Obviously, ${\bf V}^{-1}$ does not exist for comonotonic assets
and the derivation given in
appendix~\ref{app:A} does not hold, but equation~(\ref{eq:equiv_sys}) remains
well-defined and still has a unique solution $\sigma_k = (\sum w_k
\chi_k)^\frac{1}{c-1}$ which yields the scale factor given in
theorem~\ref{th:comonotony}.

\subsection{Summary}

In the previous sections, we have shown that the wealth
distribution $F_S(x)$ of a portfolio made of assets with modified Weibull
distributions with the same exponent $c$ remains equivalent in the
tail to a modified Weibull distribution ${\cal W}(c, \hat \chi)$.
Specifically, 
\be
F_S(x) \sim \lambda_- ~ F_Z(x)~,
\label{erjslas}
\ee
when $x \rightarrow - \infty$, and where $Z \sim {\cal W}(c, \hat \chi)$.
Expression (\ref{erjslas}) defines the proportionality factor or
weight $\lambda_-$ of the negative tail of the portfolio wealth distribution $F_S(x)$.
Table~\ref{table} summarizes the 
value of the scale parameter $\hat \chi$ for the different types of
dependence we have studied. In addition, we give the value of the
coefficient $\lambda_-$, which may also depend on the weights of the
assets in the portfolio in the case of dependent assets.

\begin{table}
\begin{center}
\begin{tabular}{|c|c|c|}
\hline
& $\hat \chi$ & $\lambda_-$ \\
\hline
& & \\
 & $\left(\sum_{i=1}^N |w_i \chi_i|^\frac{c}{c-1}
\right)^\frac{c-1}{c}$,  $c>1$ & $ \left[
\frac{c}{2(c-1)}\right]^\frac{N-1}{2}$\\
Independent Assets & & \\
& $\max \{ |w_1 \chi_1|,  \cdots, |w_N \chi_N| \}$,   $c \le 1$ & Card
$\{  |w_i \chi_i|= \max_j \{\ |w_j \chi_j| \} \} $\\
& & \\
\hline
& & \\
Comonotonic Assets & $\sum_{i=1}^N w_i \chi_i$ & $1$ \\
& & \\
\hline
& & \\
Gaussian copula & $\left(\sum_i w_i \chi_i
  \sigma_i \right)^\frac{c-1}{c}$ , $c>1$ & see
appendix~\ref{app:A}\\
& & \\
\hline
\end{tabular}
\end{center}
\caption{\label{table} Summary of the various scale factors obtained for
different distribution of asset returns.}
\end{table}

\section{Value-at-Risk}
\subsection{Calculation of the VaR}

We consider a portfolio made of $N$ assets with all the same exponent $c$ and
scale parameters $\chi_i$, $i \in \{1,2,\cdots,N\}$. The weight of the
i$^{th}$ asset in the portfolio is denoted by $w_i$. By definition, the
Value-at-Risk at the loss probability $\alpha$, denoted by $\Va$, is given ,
for a continuous distribution of profit and loss, by \be
\Pr\{W(\tau)-W(0) < - \Va \}=\alpha,
\ee
which can be rewritten as
\be
\Pr \left\{ S < - \frac{\Va}{W(0)} \right\}=\alpha.
\ee
In this expression, we have assumed that all the wealth is invested in risky
assets and that the risk-free interest rate equals zero, but it is easy to
reintroduce it, if necessary. It just leads to discount ${\rm VaR}_\alpha$ by
the discount factor ${1/(1+\mu_0)}$, where $\mu_0$ denotes the risk-free
interest rate.

Now, using the fact that $F_S(x) \sim \lambda_- ~ F_Z(x)$, when $x \rightarrow
- \infty$, and where $Z \sim {\cal W}(c, \hat \chi)$, we have
\be
\label{ghfla}
\frac{1}{\lambda_-}~\Pr \left\{ S < - \frac{\Va}{W(0)} \right\} \simeq 1-\Phi
\left( \sqrt{2} \left(\frac{\Va}{W(0) ~\hat \chi} \right)^{c/2} \right),
\ee
as $\Va$ goes to infinity, which allows us to obtain a closed expression for
the asymptotic Value-at-Risk with a loss probability $\alpha$:
\bea
\Va &\simeq& W(0)~ \frac{\hat \chi}{2^{1/c}} \left[\Phi^{-1}
\left(1-\frac{\alpha}{\lambda_-} \right) \right]^{2/c}, \\
& \simeq & \xi(\alpha)^{2/c} ~W(0) \cdot \hat \chi,
\eea
where the function $\Phi(\cdot)$ denotes the cumulative Normal
distribution function and
\be
\xi(\alpha) \equiv  \frac{1}{2} \Phi^{-1}\left(1-\frac{\alpha}{\lambda_-}
\right) ~. \label{jgjher}
\ee

In the case where a fraction $w_0$ of the total
wealth is invested in the risk-free asset with interest rate $\mu_0$,
the previous equation simply becomes
\be
\label{eq:VaRRF}
\Va \simeq  \xi(\alpha)^{2/c} ~(1-w_0)\cdot W(0) \cdot \hat \chi - w_0
W(0) \mu_0.
\ee
Due to the convexity of the scale parameter $\hat \chi$, the VaR is itself
convex and therefore sub-additive. Thus, for this set of distributions, the VaR
becomes coherent when the considered quantiles are sufficiently small.

The Expected-Shortfall $ES_\alpha$, which gives the average loss
beyond the VaR at probability level $\alpha$,  is also very easily
computable:
\bea
ES_\alpha &=& \frac{1}{\alpha} \int_0^\alpha {\rm VaR}_u ~du\\
&=& \zeta(\alpha)(1- w_0)\cdot W(0) \cdot \hat \chi
- w_0 W(0) \mu_0,
\eea
where $\zeta(\alpha)= \frac{1}{\alpha} \int_0^\alpha \xi(u)^{2/c}~ du~$.
Thus, the Value-at-Risk, the Expected-Shortfall and in fact any downside
risk measure involving only the far tail of the distribution of
returns are entirely controlled by the scale parameter $\hat \chi$. We
see that our set of multivariate modified Weibull distributions enjoy,
in the tail, exactly the same properties as the Gaussian
distributions, for which, all the risk measures are controlled by the
standard deviation.

\subsection{Typical recurrence time of large losses}

\begin{figure}
\begin{center}
\includegraphics[width=15cm]{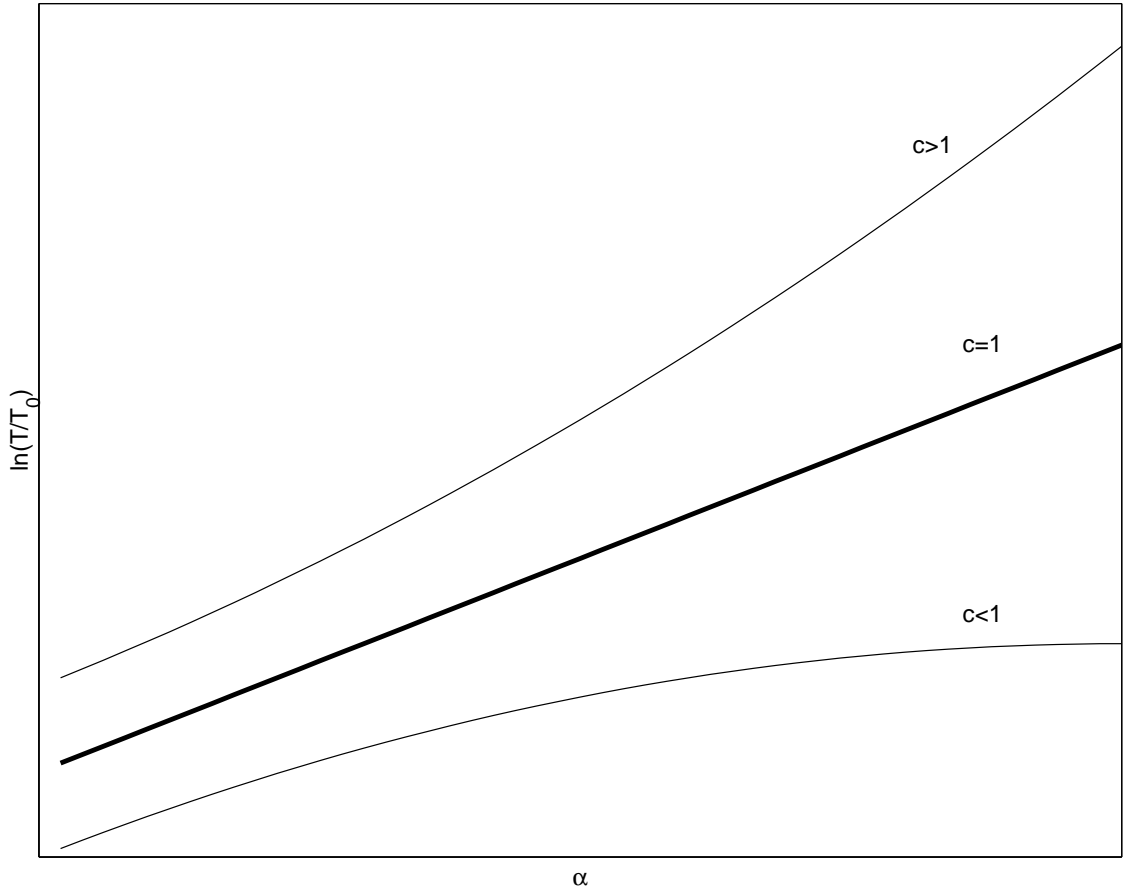}
\end{center}
\caption{\label{fig:varfreq} Logarithm $\ln \left(
\frac{T}{T_0}\right)$ of the ratio of the
recurrence time $T$ to a reference time $T_0$ for the recurrence of a
given loss $VaR$
as a function of $\beta$ defined by $\beta = \frac{\rm VaR}{{\rm VaR}^*}$.
${\rm VaR}^*$ (resp.
${\rm VaR}$) is the Value-at-Risk over a time interval $T_0$ (resp.
$T$). }
\end{figure}

Let us translate these formulas in intuitive form. For this, we
define a Value-at-Risk  $\mbox{VaR}^*$ which is such that its typical frequency
is $1/T_0$. $T_0$ is by definition the typical
recurrence time of a loss larger than $\mbox{VaR}^*$. In our present example,
we take $T_0$ equals $1$ year for example, i.e., $\mbox{VaR}^*$ is the typical
annual shock or crash.
Expression (\ref{ghfla}) then allows us to predict the recurrence
time $T$ of
a loss of amplitude VaR equal to $\beta$ times this reference value
$\mbox{VaR}^*$:
\be
\ln \left( \frac{T}{T_0} \right) \simeq  (\beta^c-1) \left(
\frac{\mbox{VaR}^*}{W(0)~ \hat \chi} \right)^c + {\cal O}(\ln \beta)~.
\ee
Figure \ref{fig:varfreq} shows $\ln \frac{T}{T_0}$ versus $\beta$.
Observe that $T$
increases all the more slowly with $\beta$, the smaller
is the exponent $c$. This
quantifies our expectation that large losses occur more frequently for
the ``wilder''
sub-exponential distributions than for super-exponential ones.

\section{Optimal portfolios}

In this section, we present our results on the problem of
the efficient portfolio allocation for asset distributed according
to modified Weibull distributions with the different
dependence structures studied in the previous sections.
We focus on the case
when all asset modified Weibull distributions
have the same exponent $c$, as it provides the richest and
more varied situation. 
When this is not the case and the
assets have different exponents $c_i$, $i=1,...,N$, 
the asymptotic tail of the portfolio return
distribution is dominated by the asset with the heaviest tail.
The largest risks of the portfolio are thus controlled by the single
most risky asset
characterized by the smallest exponent $c$. Such extreme risk cannot be
diversified away. In such a case, for a risk-averse investor, the best strategy focused
on minimizing the extreme risks consists in holding only the
asset with the thinnest tail, i.e., with
the largest exponent $c$. 

\subsection{Portfolios with minimum risk}

Let us consider first the problem
of finding the composition of the portfolio with minimum risks, where the risks
are measured by the Value-at-Risk. We consider that short sales are not
allowed, that the risk free interest rate equals zero and that all the wealth
is invested in stocks. This last condition is indeed the only interesting one since
allowing to invest in a risk-free asset would automatically give the
trivial solution in which the minimum risk portfolio is completely invested
in the risk-free asset.

The problem to solve reads:
\bea
&\Va^*=\min \Va = \xi(\alpha)^{2/c} ~ W(0) \cdot \min \hat \chi\\
&\sum_{i=1}^N w_i = 1 \\
&w_i \ge 0 ~~~~ \forall i.
\eea

In some cases (see table~\ref{table}), the prefactor $\xi(\alpha)$ defined
in (\ref{jgjher}) also
depends on the weight $w_i$'s through $\lambda_-$
defined in (\ref{erjslas}). But, its contribution remains subdominant
for the large losses. This allows to restrict the minimization to
$\hat \chi$ instead of $\xi(\alpha)^{2/c} \cdot \hat \chi$. 

\subsubsection{Case of independent assets}
\paragraph{``Super-exponential'' portfolio ($c>1$)\\ }

Consider assets distributed according to modified Weibull distributions
with the same exponent $c>1$.
The Value-at-Risk is given by
\be
\Va =  \xi(\alpha)^{2/c} ~ W(0) \cdot  \left(\sum_{i=1}^N |w_i
\chi_i|^\frac{c}{c-1} \right)^\frac{c-1}{2}~,
\ee
Introducing the Lagrange multiplier $\lambda$, the first order condition
yields
\be
\frac{\partial \hat \chi}{ \partial w_i } = \frac{\lambda}{
\xi(\alpha)~W(0)}~~~~\forall i,
\ee
and the composition of the minimal risk portfolio is
\be
{w_i}^* = \frac{\chi_i^{-c}}{ \sum_j
\chi_j^{-c}}
\ee
which satistifies the positivity of the Hessian matrix $H_{jk}= \left.
\frac{\partial^2 \hat \chi}{ \partial w_j \partial w_k} \right|_{\{w_i^*\}}$
(second order condition).

The minimal risk portfolio is such that
\be
\Va^* =  \frac{\xi(\alpha)^{2/c}~W(0)}{\left( \sum_i
\chi_j^{-c} \right)^\frac{1}{c}}~, ~~~~~
\mu^* =  \frac{\sum_i \chi_i^{-c} \mu_i }{\sum_j
\chi_j^{-c}} ~,
\ee
where $\mu_i$ is the return of asset $i$ and $\mu^*$ is the
return of the minimum risk portfolio.

\paragraph{sub-exponential portfolio ($c \leq 1$)\\ }

Consider assets distributed according to modified Weibull distributions
with the same exponent $c<1$.
The Value-at-Risk is now given by
\be
\Va =  \xi(\alpha)^{c/2} ~ W(0) \cdot \max \{ |w_1 \chi_1|,  \cdots, |w_N
\chi_N| \}. \ee
Since the weights $w_i$ are positive, the modulus appearing in the argument of the
max() function can be removed. It is easy to see
that the minimum of $\Va$ is obtained when
all the $w_i \chi_i$'s are equal, provided that the constraint $\sum w_i=1$ can
be satisfied. Indeed, let us start with the situation where
\be
w_1 \chi_1 = w_2 \chi_2 = \cdots = w_N \chi_N~.
\label{mhykr}
\ee
Let us decrease the weight $w_1$. Then, $w_1 \chi_1$ decreases with respect to the initial
maximum situation (\ref{mhykr}) but, in order to satisfy the constraint $\sum_i w_i =1$, at least
one of the other weights $w_j$, $j \ge 2$ has to increase, so that $w_j \chi_j$ increases,
leading to a maximum for the set of the $w_i \chi_i$'s greater than in the
initial situation where (\ref{mhykr}) holds. Therefore,
\be
w_i^* = \frac{A}{\chi_i}~, ~~~~\forall i,
\ee
and the constraint $\sum_i w_i=1$ yields
\be
A= \frac{1}{\sum_i \chi_i^{-1}},
\ee
and finally
\be
w_i^* =  \frac{\chi_i^{-1}}{\sum_j \chi_j^{-1}}~, ~~~~~ \Va^* =
\frac{\xi(\alpha)^{c/2}~ W(0)}{\sum_i \chi_i^{-1}}, ~~~~~~~~\mu^*  =
\frac{\sum_i \chi_i^{-1} \mu_i }{\sum_j \chi_j^{-1}} ~.\ee

The composition of the optimal portfolio is continuous in $c$ at the value
$c=1$. This is the consequence of the continuity as a function of $c$
at $c=1$ of the
scale factor $\hat \chi$ for a sum of independent variables.
In this regime $c\le1$, the Value-at-Risk increases as $c$ decreases only
through its dependence on
the prefactor $\xi(\alpha)^{2/c}$ since the scale factor
$\hat \chi$ remains constant.

\subsubsection{Case of comonotonic assets}

For comonotonic assets, the Value-at-Risk is
\be
\Va = \xi(\alpha)^{c/2}~W(0) \cdot \sum_i w_i \chi_i
\ee
which leads to a very simple linear optimization problem. Indeed,
denoting $\chi_1 = \min \{ \chi_1,$ $ \chi_2, \cdots, \chi_N \}$, we
have 
\be
\sum_i w_i \chi_i \ge \chi_1~ \sum_i w_i = \chi_1,
\ee
which proves that the composition of the optimal portfolio is $w_1^*=1, w_i^*
=0~~i \ge 2$ leading to
\be
\Va^* =   \xi(\alpha)^{c/2}~W(0)  \chi_1, ~~~~ \mu^*=\mu_1.
\ee
This result is not surprising since all assets move together. Thus, the
portfolio with minimum Value-at-Risk is obtained when only the less
risky asset, i.e., with the smallest scale factor $\chi_i$, is held.
In the case where there is a degeneracy 
in the smallest $\chi$ of order $p$ 
($\chi_1 = \chi_2 = ... = \chi_p= \min \{ \chi_1, \chi_2, \cdots,
\chi_N \}$), the optimal  choice lead to invest all the wealth in the asset  with
the larger expected return $\mu_j$, $j \in\{1, \cdots, p\}$.  However, in an
efficient market with rational agents, such an opportunity should not exist
since the same risk embodied by $\chi_1 = \chi_2 = ... = \chi_p$
should be remunerated by the same return $\mu_1=\mu_2=...=\mu_p$.

\subsubsection{Case of assets with a Gaussian copula}

In this situation, we cannot solve the problem analytically. We can only assert
that the miminization problem has a unique solution, since the function $\Va
(\{w_i\})$ is convex. In order to obtain the composition of the optimal portfolio,
we need to perform the following numerical analysis.

It is first needed to solve the set of equations
$\sum_i V_{ij}^{-1} \sigma_i^{c/2}  = w_j \chi_j \sigma_j^{1-c/2}$
or the equivalent set of equations given by (\ref{eq:equiv_sys}),
which can be performed by Newton's algorithm. Then one have the minimize the
quantity $\sum w_i \chi_i \sigma_i(\{w_i\})$. To this aim, one can use the
gradient algorithm, which requires the calculation of the derivatives of the
$\sigma_i$'s with respect to the $w_k$'s. These quantities are easily obtained 
by solving the {\it linear} set of equations
\be
\frac{c}{2} \cdot \sum_i V_{ij}^{-1} \sigma_i^{\frac{c}{2}-1}
\sigma_j^{\frac{c}{2}-1} \frac{\partial \sigma_i}{\partial w_k} + \left(
\frac{c}{2}-1 \right) w_j \chi_j \frac{1}{\sigma_j} \frac{\partial
\sigma_j}{\partial w_k} = \chi_j \cdot \delta_{jk}.
\ee
Then, the analytical solution for independent assets or comonotonic assets can
be used to initialize the minimization algorithm with respect to the
weights of the assets in the portfolio.

\subsection{VaR-efficient portfolios}

We are now interested in portfolios with minimum Value-at-risk, but with a
given expected return $\mu = \sum_i w_i \mu_i$. We will first consider
the case where, as previously, all the wealth is invested in risky
assets and we then will discuss the consequences of the introduction
of a risk-free asset in the portfolio.

\subsubsection{Portfolios without risky asset}

When the investors have to select risky assets only, they have to
solve the following minimization problem:
\bea
&\Va^*=\min \Va = \xi(\alpha) ~ W(0) \cdot \min \hat \chi\\
&\sum_{i=1}^N w_i \mu_i = \mu\\
&\sum_{i=1}^N w_i = 1 \\
&w_i \ge 0 ~~~~ \forall i.
\eea

In contrast with the research of the minimum risk portfolios where
analytical results have been derived, we need here to use numerical
methods in every situation. In the case of super-exponential portfolios,
with or without dependence between assets, the gradient method
provides a fast and easy to implement algorithm, while for
sub-exponential portfolios or portfolios made of comonotonic assets, one
has to use the simplex method since the minimization problem is then linear. 

Thus, although not as convenient to handle as analytical results, these
optimization problems remain easy to manage and fast to compute even
for large portfolios.

\subsubsection{Portfolios with risky asset}

When a risk-free asset is introduced in the portfolio, the expression
of the Value-at-Risk is given by equation (\ref{eq:VaRRF}), the
minimization problem becomes
\bea
&\Va^*=\min \xi(\alpha)^{2/c} ~(1-w_0)\cdot W(0) \cdot \hat \chi - w_0
W(0) \mu_0\\ 
&\sum_{i=1}^N w_i \mu_i = \mu\\
&\sum_{i=1}^N w_i = 1 \\
&w_i \ge 0 ~~~~ \forall i.
\eea
When the risk-free interest rate $\mu_0$ is non zero, we have to use the same
numerical methods as above to solve the problem. However, if we assume
that $\mu_0=0$, the problem becomes amenable analytically. Its
Lagrangian reads
\bea
{\cal L} &=& \xi(\alpha)^{2/c} ~(1-w_0)\cdot W(0) \cdot \hat \chi -
\lambda_1 \left( \sum_{i \neq 0}w_i \mu_i  - \mu \right) - \lambda_2
\left( \sum_{i=0}^N w_i -1 \right),\\
&=& \xi(\alpha)^{2/c} ~ \left(\sum_{j \neq 0}w_i \right)\cdot W(0)
\cdot \hat \chi - \lambda_1 \left( \sum_{i \neq 0}w_i \mu_i  - \mu
\right),
\eea
which allows us to show that the weights of the optimal portfolio are
\be
\label{eq:EF}
w_i^* =(1-w_0)\cdot \frac{\hat w_i}{\sum_{j=1}^N \hat
  w_j}~~~~~~{\rm and}~~~~~~~ \Va^*= \frac{\xi(\alpha)^{2/c}~ (1-w_0)
  \cdot W(0)}{2} \cdot \mu, 
\ee
where the $\hat w_i$'s are solution of the set of equations
\be
\label{eq:wi}
\hat \chi + \left( \sum_{i=1}^N \hat w_i \right) \frac{\partial \hat
  \chi}{\partial \hat w_i} = \mu_i~.
\ee
Expression (\ref{eq:EF}) shows that the efficient frontier is simply a straight
line and that any efficient portfolio is the sum of two
portfolios: a ``riskless portfolio'' in which a fraction $w_0$ of the
initial wealth is invested and a portfolio with the
remaining $(1-w_0)$ of the initial wealth invested in risky assets.
This provides another example of the two
funds separation theorem. A CAPM then holds,
since equation (\ref{eq:wi}) together with the market equilibrium
assumption yields the proportionality between any stock return and the
market return. However, these three
properties are rigorously established only for a zero risk-free
interest rate and
may not remain necessarily true as soon as the risk-free
interest rate becomes non zero. 

Finally, for practical purpose, the set of weights $w_i^*$'s
obtained under the assumption of zero risk-free interest rate $\mu_0$,
can be used to initialize the optimization algorithms when $\mu_0$
does not vanish.

\section{Conclusion}

The aim of this work has been to show that the key properties 
of Gaussian asset distributions of stability under convolution,
of the equivalence between all down-side riks measures, of coherence and
of simple use also hold for a general family of distributions
embodying both sub-exponential and super-exponential behaviors,
when restricted to their tail. We then used these results to 
compute the Value-at-Risk (VaR) and to obtain efficient porfolios
in the risk-return sense, where the risk is characterized by the
Value-at-Risk.
Specifically, we have studied
a family of modified Weibull distributions to
parameterize the marginal distributions of asset returns, extended to their
multivariate distribution with Gaussian copulas. The relevance 
to finance of the family of
modified Weibull distributions has been proved in both 
a context of conditional and
unconditional portfolio management. 
We have derived exact formulas for the tails of the
distribution $P(S)$ of returns $S$ of a portfolio of arbitrary composition of these
assets. We find that the tail of $P(S)$ is also 
asymptotically a modified Weibull distribution
with a characteristic 
scale $\chi$ function of the asset weights with different functional forms
depending on the super- or sub-exponential behavior of the marginals
and on the strength of the dependence between the assets.
The derivation of the portfolio
distribution has shown the asymptotic stability of this family of
distribution with the important economic consequence that any down-side risk
measure based upon the tail of the asset returns distribution are equivalent,
in so far as they all depends on the scale factor $\chi$ and keep the same
functional form whatever the number of assets in the portfolio may be.
Our analytical study of the properties of the VaR has shown the VaR
to be coherent. This
justifies the use of the VaR as a coherent risk measure for the class of modified Weibull
distributions and ensures that portfolio optimization problems are
always well-conditioned even when not fully analytically solvable.
The Value-at-Risk and the Expected-Shortfall have also
been shown to be (asymptotically)
equivalent in this framework.
In fine, using the large class of modified Weibull distributions,
we have provided a simple and fast method for
calculating large down-side risks, exemplified by the Value-at-Risk, for
assets with distributions of returns which fit quite reasonably the
empirical distributions.

\newpage

\appendix

\newpage

\section{Proof of theorem \ref{th:asws} : Tail equivalence for weighted
sums of modified Weibull variables}
\label{app:asws}

\subsection{Super-exponential case: $c>1$}

Let $X_1, X_2, \cdots, X_N$ be $N$ i.i.d random variables with
density $p(\cdot)$. Let us denote by $f(\cdot)$ and $g(\cdot)$ two
positive functions such that $p(\cdot)=g(\cdot) \cdot
e^{-f(\cdot)}$. Let $w_1, w_2, \cdots, w_N$ be $N$ real non-random 
coefficients, and  $S = \sum_{i=1}^N w_i x_i$.

Let ${\cal X} = \{ {\bf x} \in {\mathbb R}^N, \sum_{i=1}^N w_i
x_i=S\}$.  The density of the variable $S$ is given by
\be
\label{eq:ps}
P_S(S) = \int_{\cal X} d {\bf x} ~e^{-\sum_{i=1}^N \left[f(x_i)-\ln g(x_i)\right]}~,
\ee

We will assume the following conditions on the function $f$
\begin{enumerate}
\item $f(\cdot)$ is three times continuously differentiable and four
  times differentiable,
\item $f^{(2)}(x) >0$, for $|x|$ large enough,
\item $\lim_{x \rightarrow \pm \infty} \frac{f^{(3)}(x)}{(f^{(2)}(x))^2} =
0$,
\item $f^{(3)}$ is asymptotically monotonous,
\item there is a constant $\beta>1$ such that $\frac{f^{(3)}(\beta
    \cdot x)}{f^{(3)}(x)}$ remains bounded as $x$ goes to infinity,
\item $g(\cdot)$ is ultimately a monotonous function, regularly varying at
infinity with indice $\nu$. \end{enumerate}

Let us start with the demonstration of several propositions.

\begin{proposition}
\label{prop:lim}
under hypothesis 3, we have
\be
\lim_{x \rightarrow \pm \infty} |x| \cdot f''(x) =0.
\ee
$\square$
\end{proposition}

{\bf Proof}

Hypothesis 3 can be rewritten as
$\displaystyle \lim_{x \rightarrow \pm \infty} \frac{d}{dx}
\frac{1}{f^{(2)}(x)} = 0$, so that
\be
\label{eq:lim}
\forall \epsilon > 0, \exists A_\epsilon / x > A_\epsilon \Longrightarrow
\left| \frac{d}{dx} \frac{1}{f^{(2)}(x)} \right| \le \epsilon.
\ee

Now, since $f''$ is differentiable, $1/f''$ is also differentiable, and by the mean value theorem,
we have
\be
\label{eq:mean}
\left| \frac{1}{f''(x)} - \frac{1}{f''(y)} \right|  = |x-y| \cdot  \left|
\frac{d}{d\xi} \frac{1}{f''(\xi)} \right|
\ee
for some $\xi \in (x,y)$.

Choosing $x>y> A_\epsilon$, and applying equation (\ref{eq:lim}) together with
(\ref{eq:mean}) yields
\be
\left| \frac{1}{f''(x)} - \frac{1}{f''(y)} \right|  \le \epsilon \cdot |x-y|.
\ee
Now, dividing by $x$ and letting $x$ go to infinity gives
\be
\lim_{x \rightarrow \infty} \left| \frac{1}{x \cdot f''(x)} \right| \le
\epsilon,
\ee
which concludes the proof.
$\square$

\begin{proposition}
\label{prop:fprime}
Under assumption 3, we have
\be
\lim_{x \rightarrow \pm \infty} f'(x)=+ \infty.
\ee
$\square$
\end{proposition}

{\bf Proof}

According to assumption 3 and proposition \ref{prop:lim}, $\displaystyle
\lim_{x \rightarrow \pm \infty} x \cdot f''(x) = \infty$, which means
\be
\forall \alpha > 0, \exists A_\alpha / x > A_\epsilon \Longrightarrow
x \cdot f''(x) \ge \alpha.
\ee
This thus gives
\bea
\forall x \ge a_\alpha, ~~~~~~ x \cdot f''(x) \ge \alpha  &\Longleftrightarrow&
f''(x) \ge \frac{\alpha}{x}\\
&\Longrightarrow& \int_{A_\alpha}^x f''(t)~ dt \ge \alpha
\cdot \int_{A_\alpha}^x \frac{dt}{t}\\
&\Longrightarrow& f'(x) \ge \alpha \cdot \ln x - \alpha \cdot \ln A_\alpha +
f'(A_\alpha).
\eea
The right-hand-side of this last equation goes to infinity as $x$ goes to infinity, which
concludes the proof.
$\square$

\begin{proposition}
\label{prop:conv}
Under assumptions 3 and 6, the function $g(\cdot)$ satisfies
\be
\forall |h| \le \frac{C}{f''(x)},~~~~~~  \lim_{x \rightarrow \pm
  \infty} \frac{g(x+h)}{g(x)} = 1,
\ee
uniformly in $h$, for any positive constant $C$.
$\square$
\end{proposition}

{\bf Proof}
For $g$ non-decreasing,
we have
\be
\forall |h| \le \frac{C}{f''(x)},~~~~~ \frac{g\left( x \left(
      1-\frac{C}{x\cdot f''(x)} \right) \right)}{g(x)} \le
\frac{g(x+h)}{g(x)} \le \frac{g\left( x \left(1+\frac{C}{x\cdot
        f''(x)} \right) \right)}{g(x)} ~.
\ee
If $g$ is non-increasing, the same inequalities hold with the left and right
terms exchanged. Therefore, the final conclusion is easily shown to be independent 
of the monotocity property of $g$.
From assumption 3 and proposition \ref{prop:lim}, we have
\be
\forall \alpha > 0, \exists A_\alpha / x > A_\epsilon \Longrightarrow
x \cdot f''(x) \ge \alpha.
\ee
Thus, for all $x$ larger than $A_\alpha$ and all $|h| \le C/f''(x)$
\be
\frac{g\left( x \left(1-\frac{C}{\alpha} \right) \right)}{g(x)} \le
\frac{g(x+h)}{g(x)} \le \frac{g\left( x \left(1+\frac{C}{\alpha}
    \right) \right)}{g(x)}
\ee
Now, letting $x$ go to infinity, 
\be
\left(1-\frac{C}{\alpha} \right)^\nu \le \lim_{x \rightarrow \infty}
\frac{g(x+h)}{g(x)} \le \left(1+\frac{C}{\alpha}\right)^\nu,
\ee
for all $\alpha$ as large as we want, which concludes the proof.
$\square$

\begin{proposition}
\label{prop:sup}
Under assumptions 1, 3 and 4 we have, for any positive constant C:
\be
\forall |h| \le \frac{C}{f''(x)}, ~~~~~~  \lim_{x \rightarrow \pm \infty}
\frac{\sup_{\xi \in [x, x+h]} \left| f^{(3)}(\xi) \right|}{f''(x)^2} = 0.
\ee
$\square$
\end{proposition}

{\bf Proof}

Let us first remark that
\be
\frac{\sup_{\xi \in [x, x+h]} \left| f^{(3)}(\xi) \right|}{f''(x)^2}
=\frac{\sup_{\xi \in [x, x+h]} \left| f^{(3)}(\xi) \right|}{\left| f^{(3)}(x) \right|}
\cdot \frac{\left| f^{(3)}(x) \right|}{f''(x)^2}.
\ee
The rightmost factor in the right-hand-side of the equation above goes to zero as $x$ goes
to infinity by assumption 3. Therefore, we just have to show that the leftmost factor
in the right-hand-side remains bounded as $x$ goes to infinity to prove 
Proposition 4.

Applying assumption 4 according to which $f^{(3)}$ is asymptotically monotonous, we
have
\bea
\frac{\sup_{\xi \in [x, x+h]} \left| f^{(3)}(\xi) \right|}{\left| f^{(3)}(x)
\right|} &\le&
\frac{ \left| f^{(3)} \left( x  + \frac{C}{f''(x)}\right) \right|}{\left|
f^{(3)}(x) \right|}\\
&=& \frac{ \left| f^{(3)} \left( x \left( 1 + \frac{C}{x \cdot f''(x)}\right)
\right) \right|}{\left| f^{(3)}(x) \right|},\\
&\le&  \frac{ \left| f^{(3)} \left( x \left( 1 +
\frac{C}{\alpha}\right) \right) \right|}{\left| f^{(3)}(x)\right|},  
\label{eq:bound}
\eea
for every $x$ larger than some positive constant $A_\alpha$ by
assumption 3 and proposition \ref{prop:lim}.
Now, for $\alpha$ large enough, $1+\frac{C}{\alpha}$ is less than
$\beta$ (assumption 5) which shows that $\frac{\sup_{\xi \in [x, x+h]}
  \left| f^{(3)}(\xi) \right|}{\left| f^{(3)}(x) \right|}$ remains
bounded for large $x$, which conclude the proof.
$\square$

We can now show that under the assumptions stated above, the leading order
expansion of $P_S(S)$ for large $S$ and finite $N>1$ is obtained by a
generalization of Laplace's method which here amounts to remark that the set of
$x_i^*$'s that maximize the integrand in (\ref{eq:ps}) are solution of
\be
\label{eq:min}
f_i'(x_i^*)= \sigma(S) w_i~,
\ee
where $\sigma(S)$ is nothing but a Lagrange multiplier introduced to
minimize the expression $\sum_{i=1}^N f_i(x_i)$ under the constraint
$\sum_{i=1}^N w_i x_i=S$. This constraint shows that at least one $x_i$,
for instance $x_1$,
goes to infinity as $S \to \infty$.
Since $f'(x_1)$ is an increasing function by
assumption 2 which goes to infinity as $x_1 \to +\infty$
(proposition \ref{prop:fprime}), expression (\ref{eq:min})
shows that $\sigma(S)$ goes to infinity with $S$, as long as the weight
of the asset $1$ is not zero. Putting the divergence of 
$\sigma(S)$ with $S$ in expression (\ref{eq:min}) for $i=2,..., N$
ensures that each $x_i^*$ increases when $S$ 
increases and goes to infinity when $S$ goes to infinity.

Expanding $f_i(x_i)$ around $x_i^*$ yields
\be
f(x_i)=f(x_i^*)+f'(x_i^*) \cdot h_i+ \int_{x_i^*}^{x_i^* + h_i} dt
\int_{x_i^*}^t du ~ f''(u)
\ee
where the set of $h_i=x_i-x_i^*$ obey the condition
\be
\label{eq:hi}
\sum_{i=1}^N w_i h_i=0~.
\ee

Summing (\ref{eq:min}) in the presence of relation (\ref{eq:hi}), we obtain
\be
\sum_{i=1}^N f(x_i)= \sum_{i=1}^N f(x_i^*) + \sum_{i=1}^N \int_{x_i^*}^{x_i^* + h_i} dt
\int_{x_i^*}^t du ~ f''(u)~.
\ee

Thus $\exp(- \sum f(x_i))$ can be rewritten as follows :
\be
\exp \left[- \sum_{i=1}^N f(x_i) \right] = \exp \left[ \sum_{i=1}^N
  f(x_i^*)+ \sum_{i=1}^N \int_{x_i^*}^{x_i^* + h_i} dt
\int_{x_i^*}^t du ~ f''(u) \right]~.
\ee

Let us now define the compact set ${\cal A}_C = \{ {\bf h} \in {\mathbb R}^N,
\sum_{i=1}^N f''(x_i^*)^2 \cdot h_i^2 \le C^2 \}$ for any given positive
constant $C$ and the set ${\cal H} = \{ {\bf h} \in {\mathbb R}^N, \sum_{i=1}^N
w_i h_i=0 \}$. We can thus write
\bea
P_S(S) &=& \int_{\cal H} d {\bf h} ~e^{-\sum_{i=1}^N \left[f(x_i) - \ln g(x_i)\right]}~,\\
&=& \int_{{\cal A}_C \cap {\cal H}} d {\bf h} ~e^{-\sum_{i=1}^N
\left[  f(x_i) - \ln g(x_i) \right]} + \int_{\overline{{\cal A}_C} \cap {\cal H}} d
{\bf h} ~e^{-\sum_{i=1}^N \left[ f(x_i) - \ln g(x_i) \right]}~,    \label{eq:sep}
\eea
We are now going to analyze in turn these two terms in the right-hand-side of (\ref{eq:sep}).

{\bf First term of the right-hand-side of (\ref{eq:sep})}.\\
\noindent Let us start with the first term. We are going to show that
\be
\label{eq:eqi_ac}
\lim_{S \rightarrow \infty}\frac{\int_{{\cal A}_C \cap {\cal H}} d
  {\bf h} ~e^{-\sum_{i=1}^N
\int_{x_i^*}^{x_i^* + h_i} dt \int_{x_i^*}^t du ~ f''(u) - \ln
  g(x_i)}}{\frac{(2
  \pi)^{\frac{N-1}{2}}~\prod_i g(x_i^*)}
{\sqrt{\sum_{i=1}^N \frac{w_i^2 \prod_{j=1}^N
f_j''(x_j^*)}{f_i''(x_i^*)}}}}=1, ~~~~~\mbox{for some positive} ~~ C.
\ee
In order to prove this assertion, we will first consider the leftmost factor of
the right-hand-side of (\ref{eq:sep}):
 \bea
\prod g(x_i)~ e^{-\sum f(x_i)} &=& \prod g(x_i^*+h_i) ~
e^{- \sum \int_{x_i^*}^{x_i^* + h_i} dt
\int_{x_i^*}^t du ~ f''(u)} ,\\
&=& \prod g(x_i^* + h_i) ~ e^{-\frac{1}{2} \sum f''(x_i^*) h_i^2}
~
e^{- \sum \int_{x_i^*}^{x_i^* + h_i} dt
\int_{x_i^*}^t du ~ [f''(u) - f''(x_i^*)]}.
\eea
Since for all $\xi \in {\mathbb R}$, $e^{-|\xi|} \le e^{-\xi} \le e^{|\xi|}$,
we have
\be
e^{- \left| \sum \int_{x_i^*}^{x_i^* + h_i} dt
\int_{x_i^*}^t du ~ [f''(u) - f''(x_i^*)] \right|}
\le e^{- \sum \int_{x_i^*}^{x_i^* + h_i} dt \int_{x_i^*}^t du ~ [f''(u) -
f''(x_i^*)]} \le
e^{\left| \sum \int_{x_i^*}^{x_i^* + h_i} dt \int_{x_i^*}^t du ~ [f''(u) -
f''(x_i^*)] \right|},
\ee
and
\be
e^{-\sum \int_{x_i^*}^{x_i^* + h_i} dt
\int_{x_i^*}^t du ~ \left| f''(u) - f''(x_i^*) \right|}
\le e^{- \sum \int_{x_i^*}^{x_i^* + h_i} dt \int_{x_i^*}^t du ~ [f''(u) -
f''(x_i^*)]} \le
e^{\sum \int_{x_i^*}^{x_i^* + h_i} dt \int_{x_i^*}^t du ~ \left|  f''(u) -
f''(x_i^*) \right|},
\ee
since whatever the sign of $h_i$, the quantity
$\int_{x_i^*}^{x_i^* + h_i} dt \int_{x_i^*}^t du ~ \left |f''(u)-f''(x_i^*)
\right|$ remains always positive.

But, $|u-x_i^*| \le |h_i| \le \frac{C}{f''(x_i^*)}$, which leads, by
the mean value theorem and assumption 1, to
\bea
|f''(u) - f''(x_i^*)| &\le& \sup_{\xi \in (x_i^*, x_i^* + h_i)}
|f^{(3)}(\xi)| \cdot |u- x_i^*|,\\
& \le & \sup_{\xi \in (x_i^*, x_i^* + h_i)}
|f^{(3)}(\xi)|~\frac{C}{f''(x_i^*)},\\
& \le & \sup_{\xi \in {\cal G}_i}
|f^{(3)}(\xi)|~\frac{C}{f''(x_i^*)},~~{\rm where}~  {\cal
G}_i=\left[ x_i^*-\frac{C}{f''(x_i^*)}, x_i^* +\frac{C}{f''(x_i^*)} \right],
\eea
which yields
{\small
\be
0 \le \sum_i \int_{x_i^*}^{x_i^* + h_i} dt
\int_{x_i^*}^t du ~ \left |f''(u)-f''(x_i^*) \right| \le
\frac{1}{2} \sum_i  \sup_{\xi \in {\cal G}_i}
|f^{(3)}(\xi)|~\frac{C}{f''(x_i^*)} h_i^2~.
\ee}
Thus
\be
e^{-\frac{1}{2} \sum_i  \sup
|f^{(3)}(\xi)|~\frac{C}{f''(x_i^*)} h_i^2}
\le e^{- \sum \int_{x_i^*}^{x_i^* + h_i} dt \int_{x_i^*}^t du ~ [f''(u) -
f''(x_i^*)]} \le
e^{\frac{1}{2} \sum_i  \sup
|f^{(3)}(\xi)|~\frac{C}{f''(x_i^*)} h_i^2},
\ee
where $\sup_{\xi \in {\cal G}_i}
|f^{(3)}(\xi)|$, have been denoted by  $\sup|f^{(3)}(\xi)|$ in the previous
expression, in order not to cumber the notations.

By proposition \ref{prop:sup}, we know that for all ${\bf h} \in {\cal A}_C$
and all $\epsilon_i >0$
\be
\left| \frac{\sup |f^{(3)}(\xi)}{f''(x_i^*)} \right| \le \epsilon_i,~~~~ {\rm
for}~ x_i^*~ {\rm large~ enough,}
\ee
so that
\be
\forall \epsilon' >0~ {\rm and}~ \forall {\bf h} \in {\cal A}_C,~~~~
e^{-\frac{C \cdot \epsilon'}{2} \sum_i  h_i^2}
\le e^{- \sum \int_{x_i^*}^{x_i^* + h_i} dt \int_{x_i^*}^t du ~ [f''(u) -
f''(x_i^*)]} \le
e^{\frac{C \cdot \epsilon'}{2} \sum_i  h_i^2},
\ee
for $|{\bf x}|$ large enough.

Moreover, from proposition~\ref{prop:conv}, we have for all $\epsilon_i >0$
and $x_i^*$ large enough:
\be
\forall {\bf h} \in {\cal A}_C,~~~ (1-\epsilon_i)^\nu \le
\frac{g(x_i^*+hi)}{g(x_i^*)} \le (1+\epsilon_i)^\nu,
\ee
so, for all $\epsilon''>0$
\be
\forall {\bf h} \in {\cal A}_C,~~~ (1-\epsilon'')^{N \nu} \le
\prod_i \frac{g(x_i^*+hi)}{g(x_i^*)} \le (1+\epsilon'')^{N \nu}.
\ee
Then for all $\epsilon>0$ and $|{\bf x}|$ large enough, this yields :
\be
(1-\epsilon)^{N \nu}~ e^{-\frac{1}{2} \sum_i (f''(x_i^*)+ C \cdot
\epsilon)\cdot h_i^2} \le \prod_i \frac{g(x_i)}{g(x_i^*)} e^{-\sum
f(x_i)} \le (1+ \epsilon)^{N \nu}~ e^{-\frac{1}{2} \sum_i (f''(x_i^*)- C
\cdot \epsilon)\cdot h_i^2},
\ee
for all ${\bf h} \in {\cal A}_C$. Thus, integrating over all the ${\bf h} \in
{\cal A}_C \cap {\cal H}$  and by continuity of the mapping
\be
G( {\bf Y}) = \int_{{\cal A}_C \cap {\cal H}} d {\bf h} ~ g({\bf h},
{\bf Y})
\ee
where  $g({\bf h}, {\bf Y}) = e^{- \frac{1}{2}\sum  Y_i \cdot   h_i^2}$, we can
conclude that,
\be
\frac{\int_{{\cal A}_C \cap {\cal
H}} \prod g(x_i)~ e^{-\sum f(x_i)}}{\prod g(x_i^*)~ \int_{{\cal A}_C \cap
{\cal H}} d {\bf h} ~e^{- \frac{1}{2}\sum f''(x_i^*) h_i^2}}  \xrightarrow[]{S
\rightarrow \infty} 1.
\ee
Now, we remark that
\be
 \int_ {\cal
H} d {\bf h} ~e^{- \frac{1}{2}\sum f''(x_i^*) h_i^2} = \int_{{\cal A}_C \cap {\cal
H}} d {\bf h} ~e^{- \frac{1}{2}\sum f''(x_i^*) h_i^2} + \int_{\overline{{\cal
A}_C} \cap {\cal H}} d {\bf h} ~e^{- \frac{1}{2}\sum f''(x_i^*) h_i^2}  ,
\ee
with
\be
\int_{{\cal H}} d {\bf h} ~e^{- \frac{1}{2}\sum
  f''(x_i^*) h_i^2} =  \frac{(2 \pi)^{\frac{N-1}{2}}}
{\sqrt{\sum_{i=1}^N \frac{w_i^2 \prod_{j=1}^N
f_j''(x_j^*)}{f_i''(x_i^*)}}},
\ee
and
\be
 \int_{\overline{{\cal
A}_C} \cap {\cal H}} d {\bf h} ~e^{- \frac{1}{2}\sum f''(x_i^*) h_i^2} \sim
{\cal O} \left( e^{- \frac{\alpha}{f''(x^*)} } \right), ~~~~~ \alpha >0,
\ee
where $x^* = \max \{ x_i^*\}$ (note that $1/f''(x) \to \infty$
with $x$ by Proposition 1). Indeed, we clearly have
\bea
 \int_{\overline{{\cal A}_C} \cap {\cal H}} d {\bf h} ~e^{- \frac{1}{2}\sum
f''(x_i^*) h_i^2} &\le&  \int_{\overline{{\cal A}_C}} d {\bf h} ~e^{- \frac{1}{2}\sum
f''(x_i^*) h_i^2},\\
&=& \frac{(2\pi)^{N/2}}{\sqrt{\prod f''(x_i^*)}}  ~ \int_{\overline{{\cal
B}_C}} d{\bf u} ~ \prod_i \frac{e^{-\frac{1}{2}
\frac{u_i^2}{f''(x_i^*)}}}{\sqrt{2\pi~ f''(x_i^*)}}~,
\label{gjsllw}
\eea
where we have performed the change of variable $u_i = f''(x_i^*) \cdot h_i$ and
denoted by ${\cal B}_C$ the set $\{ {\bf h} \in {\mathbb R}^N, \sum u_i^2 \le
C^2 \}$. Now, let $x^*_{\rm max} = \max \{x_i^*\}$ and
$x^*_{\rm min} = \min \{x_i^*\}$. Expression (\ref{gjsllw}) then gives
\bea
 \int_{\overline{{\cal A}_C} \cap {\cal H}} d {\bf h} ~e^{- \frac{1}{2}\sum
f''(x_i^*) h_i^2} &\le& \frac{(2\pi)^{N/2}}{f''(x^*_{\rm min})^{N/2}}  ~
\int_{\overline{{\cal B}_C}} d{\bf u}  ~ \frac{e^{-\frac{1}{2~
f''(x^*_{\rm max})} \sum u_i^2}}{(2\pi~ f''(x^*_{\rm min}))^{N/2}},\\
&=&  {\cal S}_{N-1}~
 \frac{f''(x^*_{\rm max})^{N/2}}{ f''(x^*_{\rm min})^N}~
\Gamma \left( \frac{N}{2}, \frac{C^2}{2~ f''(x^*_{\rm max})} \right),\\
&\simeq&  {\cal S}_{N-1}~
\frac{f''(x^*_{\rm max})^{N/2}}{f''(x^*_{\rm min})^N}~\left(
\frac{C^2}{2~ f''(x^*_{\rm max})} \right)^{\frac{N}{2}-1}
\cdot e^{- \frac{C^2}{2~ f''(x^*_{\rm max})}},
\eea
which decays exponentially fast for large $S$ (or large $x^*_{\rm max}$) as long as $f''$ goes
to zero at infinity, i.e, for any function $f$ which goes to infinity
not faster than $x^2$.
So, finally
\be
\int_{{\cal A}_C \cap {\cal
H}} d {\bf h} ~e^{- \frac{1}{2}\sum f''(x_i^*) h_i^2}  = \frac{(2 \pi)^{\frac{N-1}{2}}}
{\sqrt{\sum_{i=1}^N \frac{w_i^2 \prod_{j=1}^N f_j''(x_j^*)}{f_i''(x_i^*)}}}
+  {\cal  O} \left( e^{- \frac{\alpha}{f''(x^*_{\rm max})}} \right)~,
\ee
which concludes the proof of equation (\ref{eq:eqi_ac}).

{\bf Second term of the right-hand-side of (\ref{eq:sep})}.\\
\noindent We now have to show that
\be
\int_{\overline{{\cal A}_C} \cap {\cal H}} d {\bf h} ~
e^{- \sum f(x_i^*+h_i) - g(x_i^* + h_i)}
\ee
can be neglected. This is obvious since, by assumption 2 and 6, the
function $f(x) - \ln g(x)$ remains convex for $x$ large enough, which
ensures that $f(x) - \ln g(x) \ge C_1 |x|$ for some positive constant
$C_1$ and $x$ large enough. Thus, choosing the constant $C$ in ${\cal
  A}_C$ large enough, we have
\be
\int_{\overline{{\cal A}_C} \cap {\cal H}} d{\bf h} ~e^{-\sum_{i=1}^N
  f(x_i) - \ln g(x_i)} \le \int_{\overline{{\cal A}_C} \cap {\cal H}} d
{\bf h} ~e^{-C_1\sum_{i=1}^N  |x_i^* +h_i|} \sim 
{\cal O}\left(e^{-\frac{\alpha'}{f''(x^*)}} \right)~.
\ee
Thus, for $S$ large enough, the density $P_S(S)$ is asymptotically
equal to
\be
P_S(S)=\prod_i g(x_i^*)~
\frac{(2 \pi)^{\frac{N-1}{2}}}{\sqrt{\sum_{i=1}^N \frac{w_i^2 \prod_{j=1}^N
f_j''(x_j^*)}{f_i''(x_i^*)}}}~.
\ee

In the case of the modified Weibull variables, we have
\be
f(x)= \left( \frac{|x|}{\chi} \right)^c,
\ee
and
\be
g(x)= \frac{c}{2 \sqrt{\pi} \chi^{c/2}} \cdot |x|^{\frac{c}{2}-1},
\ee
which satisfy our assumptions if and only if $c>1$. In such a
case, we obtain
\be
x_i^* = \frac{w_i^\frac{1}{c-1}}{\sum w_i^\frac{c}{c-1}}\cdot S,
\ee
which, after some simple algebraic manipulations, yield
\be
P(S) \sim \left[ \frac{c}{2(c-1)} \right]^\frac{N-1}{2} \frac{c}{2
 \sqrt{\pi}} \frac{1}{\hat \chi^{c/2}}|S|^{\frac{c}{2}-1} e^{-\left(
   \frac{|S|}{\hat \chi} \right)^c}
\ee
with
\be
\hat \chi = \left(\sum w_i^\frac{c}{c-1}\right)^\frac{c-1}{c} \cdot \chi.
\ee
as announced in theorem \ref{th:asws}.

\subsection{Sub-exponential case:  $c \le 1$}

Let $X_1, X_2, \cdots, X_N$ be $N$ i.i.d sub-exponential  modified Weibull
random variables ${\cal W}(c, \chi)$, with distribution function $F$. Let us
denote by $G_S$ the distribution function of the variable
\be
S_N = w_1 X_1 + w_2 X_2 + \cdots + w_N X_N,
\ee
where $w_1, w_2, \cdots, w_N$ are real non-random coefficients.

Let $w^* = \max\{|w_1|, |w_2|, \cdots , |w_N| \}$. Then, theorem 5.5 (b) of
\citeasnoun{GK98}  states that
\be
\lim_{x \rightarrow \infty} \frac{G_S(x/w^*)}{F(x)}= \mbox{Card}~ \{ i \in
\{1,2, ,N\} :  |w_i|=w^* \}.
\ee
By definition \ref{def:te}, this allows us to conclude that $S_N$ is
equivalent in the upper tail to $Z \sim {\cal W}(c, w^* \chi)$.

A similar calculation yields an analogous result for the lower tail.

\newpage

\section{Asymptotic distribution of the sum of Weibull variables with
a Gaussian copula.}
\label{app:A}

We assume that the marginal distributions are given by the modified
Weibull distributions:
\be
P_i(x_i)=\frac{1}{2 \sqrt{\pi}}\frac{c}{\chi_i^{c/2}} |x_i|^{c/2-1}
e^{-\left(\frac{|x_i|}{\chi_i} \right)^c}
\ee
and that the $\chi_i$'s are all equal to one, in order not to cumber
the notation. As in the proof of corollary \ref{co:1}, it will be sufficient to
replace $w_i$ by $w_i \chi_i$ to reintroduce the scale factors.

Under the Gaussian copula assumption, we obtain the following form
for the multivariate distribution :
\be
P(x_1, \cdots , x_N) = \frac{c^N}{2^N \pi^{N/2}\sqrt{\det V}} \prod_{i=1}^N
x_i^{c/2-1} \exp \left[-\sum_{i,j}
V_{ij}^{-1}x_i^{c/2}x_j^{c/2} \right]~.
\ee
Let
\be
f(x_1, \cdots, x_N) = \sum_{i,j}
V_{ij}^{-1}{x_i}^{c/2}{x_j}^{c/2}~.
\ee

We have to minimize $f$ under the constraint $\sum w_i x_i = S$. As
for the independent case, we introduce a Lagrange multiplier $\lambda$
which leads to
\be
c\sum_{j}
V_{jk}^{-1}{x_j^*}^{c/2}{x_k^*}^{c/2-1}= \lambda w_k~.
\ee
The left-hand-side of this equation is a homogeneous function of degree $c-1$ in
the $x_i^*$'s, thus necessarily
\be
x_i^* =
\left(\frac{\lambda}{c}\right)^\frac{1}{c-1}\cdot
\sigma_i,
\ee
where the $\sigma_i$'s are solution of
\be
\label{eq:154}
\sum_{j} V_{jk}^{-1}{\sigma_j}^{c/2}{\sigma_k}^{c/2-1}=w_k~.
\ee
The set of equations (\ref{eq:154}) has
a unique solution due to the convexity of the
minimization problem. This set of equations can be
easily solved by a numerical method like Newton's algorithm. It
is convenient to simplify the problem and avoid the inversion of the matrix
$V$, by rewritting (\ref{eq:154}) as
\be
\sum_{k} V_{jk}~w_k~{\sigma_k}^{1-c/2} = \sigma_j^{c/2}~.
\ee

Using the constraint $\sum w_i x_i^* =S$,
we obtain
\be
\left(\frac{\lambda}{c}\right)^\frac{1}{c-1} =
  \frac{S}{\sum w_i \sigma_i},
\ee
so that
\be
x_i^* = \frac{\sigma_i}{\sum w_i \sigma_i}\cdot S.
\ee
Thus
\bea
f(x_1, \cdots , x_N) &=& f(x_1^*, \cdots , x_N^*) + \sum_i
\frac{\partial f}{\partial x_i} h_i + \frac{1}{2} \sum_{ij} \frac{\partial^2
f}{\partial x_i \partial x_j} h_i h_j + \cdots \\
&=& \frac{S^c}{(\sum w_i \sigma_i)^{c-1}} +
\frac{1}{2} \sum_{ij} \frac{\partial^2
f}{\partial x_i \partial x_j} h_i h_j + \cdots ~,
\eea
where, as in the previous section, $h_i = x_i-x_i^*$ and the
derivatives of $f$ are expressed at $x_1^*, ..., x_N^*$.

It is easy to check that the $n^{th}$-order derivative of $f$ with respect
to the $x_i$'s evaluated at $\{x_i^*\}$ is proportional to $S^{c-n}$.
In the sequel, we
will use the following notation~:
\be
\left. \frac{\partial^n f}{\partial x_{i_1} \cdots \partial x_{i_n}}
\right|_{\{x_i^*\}} = M_{{i_1} \cdots {i_n}}^{(n)} S^{c-n}~.
\ee

We can write :
\be
f(x_1, \cdots , x_N) =  \frac{S^c}{(\sum w_i \sigma_i)^{c-1}} +
\frac{S^{c-2}}{2} \sum_{ij} M_{ij}^{(2)} h_i h_j 
+ \frac{S^{c-3}}{6} \sum_{ijk} M_{ijk}^{(3)} h_i h_j h_k + \cdots
\ee
up to the fourth order. This leads to
\bea
P(S)\propto e^{-  \frac{S^c}{(\sum w_i \sigma_i)^{c-1}}} \int dh_1
\cdots dh_N e^{- \frac{S^{c-2}}{2} 
\sum_{ij} M_{ij}^{(2)} h_i h_j }~\delta \left( \sum w_i h_i \right)
\times \nonumber \\
\times \left[ 1 + \frac{S^{c-3}}{6} \sum_{ijk} M_{ijk}^{(3)} h_i h_j h_k +
\cdots \right]~.
\eea

Using the relation $\delta\left(\sum w_i h_i \right) = \int
\frac{dk}{2 \pi}~e^{-ik \sum_j w_j h_j}$, we obtain :
\bea
P(S)\propto e^{- \frac{S^c}{(\sum w_i \sigma_i)^{c-1}}} \int
\frac{dk}{2 \pi} \int dh_1 \cdots dh_N e^{- 
\frac{S^{c-2}}{2}
\sum_{ij} M_{ij}^{(2)} h_i h_j -ik \sum_j w_j h_j}
\times \nonumber \\
\times \left[ 1 + \frac{S^{c-3}}{6} \sum_{ijk} M_{ijk}^{(3)} h_i h_j h_k +
\cdots \right]~,
\eea
or in vectorial notation :
\bea
P(S)\propto e^{-\frac{S^c}{(\sum w_i \sigma_i)^{c-1}}}  \int
\frac{dk}{2 \pi} \int d{\bf h}~ e^{- \frac{S^{c-2}}{2} 
{\bf h^t M^{(2)} h} -ik {\bf w^t h}}
\times \nonumber \\
\times \left[ 1 + \frac{S^{c-3}}{6} \sum_{ijk} M_{ijk}^{(3)} h_i h_j h_k +
\cdots \right]~.
\eea

Let us perform the following standard change of variables~:
\be
{\bf h} = {\bf h'}-\frac{ik}{S^{c-2}} {\bf {M^{(2)}}^{-1} w}~.
\ee
(${\bf {M^{(2)}}^{-1}}$ exists since $f$ is assumed convex and thus
${\bf M^{(2)}}$ positive)~:
\be
\frac{S^{c-2}}{2}{\bf h^t M^{(2)} h} +ik {\bf w^t h} =
\frac{S^{c-2}}{2}{\bf h'^t M^{(2)} h'} + \frac{k^2}{2S^{c-2}} {\bf
w^t  {M^{(2)}}^{-1} w}~.
\ee
This yields
\bea
P(S)\propto e^{- \frac{S^c}{(\sum w_i \sigma_i)^{c-1}}} \int
\frac{dk}{2 \pi}~e^{-\frac{k^2}{2S^{c-2}} {\bf 
w^t  {M^{(2)}}^{-1} w}}\times \nonumber \\
\times \int d{\bf h}~ e^{-\frac{S^{c-2}}{2}
\left({\bf h}+\frac{ik}{S^{c-2}} {\bf {M^{(2)}}^{-1} w} \right)^t
 M^{(2)}\left({\bf h}+\frac{ik}{S^{c-2}} {\bf {M^{(2)}}^{-1} w} \right) }
\left[ 1 + \frac{S^{c-3}}{6} \sum_{ijk} M_{ijk}^{(3)} h_i h_j h_k +
\cdots \right]~.
\eea

Denoting by $\langle \cdot \rangle_h$ the average with respect
to the Gaussian distribution of ${\bf h}$ and by $\langle \cdot
\rangle_k$ the average with respect to the Gaussian distribution of
$k$, we have~:
\bea
P(S)\propto \sqrt{ \frac{\det {\bf{M^ {(2)} }^{-1}} } { {\bf w^t {M^
{(2)} }^{-1} w} } }~
(2 \pi S^{2-c})^{\frac{N-1}{2}}~
e^{- \frac{S^c}{(\sum w_i \sigma_i)^{c-1}}} \times
\nonumber \\
\times \left[ 1 + \frac{S^{c-3}}{6} \sum_{ijk} M_{ijk}^{(3)}\langle \langle
h_i h_j h_k \rangle_{\bf h} \rangle_k +  \frac{S^{c-4}}{24}
\sum_{ijkl}  M_{ijkl}^{(4)}\langle \langle
h_i h_j h_k h_l\rangle_{\bf h} \rangle_k + \cdots \right]~.
\eea
We now invoke Wick's theorem \footnote{See for instance 
\cite{brezin} for a general introduction, \cite{Sornettepre} for an early application
to the portfolio problem and \cite{Sornette1}
for a systematic utilization with the help of diagrams.}, which states that each term $\langle
\langle h_i \cdots
h_p \rangle_{\bf h} \rangle_k$ can be expressed as a product of pairwise
correlation coefficients. Evaluating the average with
respect to the symmetric distribution of $k$, it is obvious that
odd-order terms will vanish and that the count of powers of $S$ involved in
each even-order term shows that all are sub-dominant. So, up to the leading
order~:
\be
\label{eq:ohk}
P(S)\propto \sqrt{ \frac{\det {\bf{M^ {(2)} }^{-1}} } { {\bf w^t {M^
{(2)} }^{-1} w} } }~
(2 \pi S^{2-c})^{\frac{N-1}{2}}~
e^{- \frac{S^c}{(\sum w_i \sigma_i)^{c-1}}}~.
\ee

The matrix $M^{(2)}$ can be calculated, which yields
\bea
M^{(2)}_{kl} &=& \frac{1}{(\sum w_i \sigma_i)^{c-2}} \left[ c \left(
    \frac{c}{2}-1 \right) \frac{w_k}{\sigma_k}~\delta_{kl} +
\frac{c^2}{2} V_{kl}^{-1} \sigma_k^{\frac{c}{2}-1}~ \sigma_l^{\frac{c}{2}-1}
 \right],\\
&=& \frac{1}{(\sum w_i \sigma_i)^{c-2}}~ \tilde M_{kl},
\eea
and shows that
\be
\label{eq:tyf}
\sqrt{ \frac{\det {\bf{M^ {(2)} }^{-1}} } { {\bf w^t {M^{(2)} }^{-1}
w}}} = \left(\sum w_i \sigma_i \right)^{(N-1)\left(\frac{c}{2}-1\right)}
\sqrt{ \frac{\det {\bf{\tilde M }^{-1}} } { {\bf w^t
      {\tilde M}^{-1} w}}}.
\ee

The inverse matrix $\tilde M^{-1}$ satisfies  $\sum _l \tilde M_{kl} \cdot
(\tilde M^{-1})_{lj} = \delta_{kj}$ which can be rewritten:
\be
c \left(\frac{c}{2}-1 \right) \frac{w_k}{\sigma_k}~(\tilde M^{-1})_{kj} +
\frac{c^2}{2} \sum_l V_{kl}^{-1} \cdot (\tilde M^{-1})_{lj}
\sigma_k^{\frac{c}{2}-1}~ \sigma_l^{\frac{c}{2}-1} = \delta_{kj}
\ee
or equivalently
\be
c \left(\frac{c}{2}-1 \right) w_k~(\tilde M^{-1})_{kj} +
\frac{c^2}{2} \sum_l V_{kl}^{-1} \cdot (\tilde M^{-1})_{lj}
\sigma_k^{\frac{c}{2}}~ \sigma_l^{\frac{c}{2}-1} = \delta_{kj} \cdot \sigma_k
\ee
which gives
\be
c \left(\frac{c}{2}-1 \right) \sum_{j,k}w_k~(\tilde M^{-1})_{kj}~w_j +
\frac{c^2}{2} \sum_{j,k,l} V_{kl}^{-1} \cdot (\tilde M^{-1})_{lj}
\sigma_k^{\frac{c}{2}}~ \sigma_l^{\frac{c}{2}-1} w_j= \sum_{j,k} \delta_{kj}
\cdot \sigma_k w_j ~. \ee
Summing the rightmost factor of the left-hand-side over $k$, and accounting for equation
(\ref{eq:154}) leads to \be
c \left(\frac{c}{2}-1 \right) \sum_{j,k}w_k~(\tilde M^{-1})_{kj}~w_j +
\frac{c^2}{2} \sum_{j,l} w_l ~ (\tilde M^{-1})_{lj}  w_j= \sum_j \sigma_j w_j
~. \ee
so that
\be
\label{eq:mju}
{\bf w^t \tilde M^{-1} w} = \frac{1}{c(c -1)} \sum_j w_j \sigma_j    ~.
\ee
Moreover
\be
\label{eq:aer}
\frac{c^N}{2^N \pi^{N/2}\sqrt{\det V}} \prod_{i=1}^N
{x_i^*}^{c/2-1} =  \frac{c^N}{2^N \pi^{N/2}\sqrt{\det
    V}} \frac{\prod \sigma_i^{\frac{c}{2}-1}}{\left( \sum
    w_i \sigma_i \right)^{N \left(\frac{c}{2}-1\right)}}~ S^{N
  \left(\frac{c}{2}-1\right)}~.
\ee
Thus, putting together equations (\ref{eq:ohk}), (\ref{eq:tyf}),
(\ref{eq:mju}) and (\ref{eq:aer}) yields
\be
P(S) \simeq \sqrt{c(c-1)~\frac{\det \tilde M^{-1}}{\det V}} \frac{c^{N-1} \prod
\sigma_i^{c/2-1}}{2^{(N-1)/2}} \cdot
\frac{1}{2 \sqrt{\pi}}\frac{c}{\hat \chi^{c/2}} |S|^{c/2-1}
e^{-\left(\frac{|S|}{\hat \chi} \right)^c},
\ee
with
\be
\hat \chi = \left( \sum w_i \chi_i \sigma_i \right)^\frac{c-1}{c}.
\ee

\newpage

\clearpage

\end{document}